\documentclass{article}

\usepackage[numbers]{natbib}
\usepackage{PRIMEarxiv}
\usepackage{float}
\usepackage[utf8]{inputenc} 
\usepackage[T1]{fontenc}    
\usepackage{hyperref}       
\usepackage{url}            
\usepackage{booktabs}       
\usepackage{amsfonts}       
\usepackage{nicefrac}       
\usepackage{microtype}      
\usepackage{lipsum}
\usepackage{fancyhdr}       
\usepackage{graphicx}       
\graphicspath{{media/}}     
\usepackage[ruled,vlined]{algorithm2e}

\usepackage{color}
\usepackage{tikz}

\newcommand{\probb}{\text{I\kern-0.15em P}}
\usepackage{caption}
\usepackage{subcaption}
\usepackage{changepage}
\usepackage{fancyvrb}
\usepackage{amsthm}
\usepackage{xcolor}
\usepackage{amsfonts}
\newcommand{\norm}[1]   {\left\| #1 \right\|}

\usetikzlibrary{shapes,decorations,arrows,calc,arrows.meta,fit,positioning}

\newtheorem{definition}{Definition}

\newtheorem{theorem}{Theorem}

\theoremstyle{definition}

\newtheorem*{remark}{Remark}

\usepackage[utf8]{inputenc}
\usepackage[T1]{fontenc}
\usepackage{amsmath}
\usepackage{amsfonts}
\usepackage{amssymb}
\usepackage{graphicx}
\usepackage{stackengine}
\usepackage{scalerel}
\usepackage{authblk}

\usepackage[utf8]{inputenc}
\usepackage[T1]{fontenc}
\usepackage{moreverb,url}

\usepackage{amsmath}
\usepackage{amsfonts}
\usepackage{amssymb}
\usepackage{graphicx}
\usepackage{hyperref}
\usepackage{moreverb,url,multirow, tabularx}

\newtheorem{proposition}[theorem]{Proposition}
\newtheorem{corollary}[theorem]{Corollary}

\newtheorem{assumption}{Assumption}
\allowdisplaybreaks

\usepackage{pgfplotstable}
\usepackage{booktabs}
\usepackage{xspace}
\newcommand\BibTeX{{\rmfamily B\kern-.05em \textsc{i\kern-.025em b}\kern-.08em
		T\kern-.1667em\lower.7ex\hbox{E}\kern-.125emX}}

\DeclareSymbolFont{bbold}{U}{bbold}{m}{n}
\DeclareSymbolFontAlphabet{\mathbbold}{bbold}

\def\spacingset#1{\renewcommand{\baselinestretch}%
{#1}\small\normalsize} \spacingset{1}

 \title{Conformal uncertainty quantification using kernel depth measures in separable Hilbert spaces}
 \author{Marcos Matabuena \thanks{\url{mmatabuena@hsph.harvard.edu}
    } \\
    Department of Biostatistics, Harvard University
    \and
    Rahul Ghosal \\
    Department of Epidemiology and Biostatistics, University of South Carolina \\
    \and  Pavlo Mozharovskyi\\
    LTCI, Telecom Paris, Institut Polytechnique de Paris \\ \and Oscar Hernan Madrid Padilla \\ UCLA
\and  Jukka-Pekka Onnela \\
    Department of Biostatistics, Harvard University    
  }  

\begin{document}
\maketitle

\begin{abstract}
Depth measures have gained popularity in the statistical literature for defining level sets in complex data structures like multivariate data, functional data, and graphs. Despite their versatility, integrating depth measures into regression modeling for establishing prediction regions remains underexplored. To address this gap, we propose a novel method utilizing a model-free uncertainty quantification algorithm based on conditional depth measures and conditional kernel mean embeddings. This enables the creation of tailored prediction and tolerance regions in regression models handling complex statistical responses and predictors in separable Hilbert spaces. Our focus in this paper is exclusively on examples where the response is a functional data object. To enhance practicality, we introduce a conformal prediction algorithm, providing non-asymptotic guarantees in the derived prediction region. Additionally, we establish both conditional and unconditional consistency results and fast convergence rates in some special homoscedastic cases. We evaluate the model finite sample performance in  extensive simulation studies with different function objects as probability distributions and  functional data. Finally, we apply the approach in a digital health application related to physical activity, aiming to offer personalized recommendations in the US. population  based on individuals'  characteristics.
\end{abstract}


\section{Introduction}
The study and exploration of  depth measures  \cite{zuo2000general, briend2023quality},
 for different statistical modeling tasks such as exploratory analysis \cite{liu1999multivariate}
and classification \cite{li2012dd} have attracted significant attention in the statistical literature, see, e.g.  \cite{LopezPintadoR09,NarisettyN16,10.1093/jssam/smz060}, to name but a few, see also \cite{MoslerM22} for a recent survey. Depth serves as a valuable alternative to traditional centrality measures, like the mean or median, establishing a notion of order statistics particularly useful for multivariate data and high-dimensional objects, such as functional data, where a natural order is absent.

The concept of depth was first introduced by Mahalanobis and Tukey (see for example \cite{mclachlan1999mahalanobis, berrendero2020mahalanobis, tukey1975mathematics}), with a important historical background in bivariate rank test \cite{hodges1955bivariate}.  Donoho performed a
 subsequent theoretical examination of their properties, especially robustness \cite{donoho1982breakdown,donoho1992breakdown}.

 The objective is to define a rank using a  sequence of random variables \(Y_{1}, \cdots, Y_{n}\) within a set \(\mathcal{Y}\), assessing the centrality of new data points \(y\) in relation to the distribution of \(Y,\) denoted \(P\). Early depth measures ranked points by their Euclidean distance to the data distribution's geometric center, similar to the multivariate mean or median, as exemplified by the Mahalanobis distance. This approach, however has limitations with non-Gaussian distributions, such as multimodal and asymmetrical distributions, where a simple geometrical distance does not capture the distribution's complexity. These challenges prompted a second wave of depth research in the 1980s and 1990s, introducing new depth measures like Oja, simplicial, and majority depths (see \cite{liu1990notion,oja1983descriptive,singh1991notion,liu1999multivariate, zuo2000general}
  for a comprehensive review).

Depth measures gained prominence in the 1970s for exploratory analysis, with their application now extending to outlier detection \cite{chen2008outlier}, variable selection   \cite{RePEc:spr:alstar:v:105:y:2021:i:2:d:10.1007_s10182-021-00391-y}, hypothesis testing \cite{singh2022some, liu1999multivariate}, clustering \cite{pandolfo2023clustering}, missing data \cite{doi:10.1080/01621459.2018.1543123}
 and classification \cite{li2012dd}.
The robustness of depth measures, particularly their role in robust statistics \citep{denecke2012consistency} in terms of influence functions \citep{dang2009influence}
 has been a significant area of theoretical analysis.

A third wave in depth measure research is currently underway, aiming to define general depth measures applicable to a variety of statistical objects, including functional, high-dimensional, and non-Euclidean data types such as graphs and multivariate densities in metric spaces \cite{virta2023spatial,geenens2023statistical, dai2023tukey} or spatial curves \cite{LafayeDeMicheauxMV20}.
  These advances are particularly relevant for modern medical and public health applications \cite{rodriguez2022contributions,lugosi2024uncertainty},
 facilitating deep patient phenotyping and the analysis of complex data structures like biosensor signals and medical images. In diabetes research  \cite{matabuena2021glucodensities}, for instance, individual continuously monitored glucose levels maybe to used to  identify anomalous low and high-glucose events (clinically corresponding to hypo- and hyperglycemia ranges).

 In a seminal work, \cite{li2008multivariate} established level sets and tolerance regions based on depth measures from an unconditional perspective.  Tolerance regions   \citep{murphy1948non,  krishnamoorthy2009statistical} are a relevant data analysis tool in various fields, particularly in quality control and medical science for establishing normal ranges for clinical biomarkers derived from healthy populations. These regions facilitate the reclassification of patients, into distinct categories, such as healthy and disease group \citep{young2020nonparametric}.

Depth measures have been shown to outperform traditional non-parametric kernel density estimation methods in reconstructing the level sets 
 of data distributions across different conditions and data structures \citep{tsybakov1997nonparametric}, in including functional spaces \citep{LopezPintadoR09}. 
 Notably, depth measures, through depth bands, exhibit greater robustness against the curse of dimensionality, are robust to outliers, and offer improved convergence rates in many scenarios (see for example
 \cite{pmlr-v70-jiang17b} and \cite{zhang2002some}).
Conformal inference, introduced by \cite{shafer2008tutorial}, is designed to quantify uncertainty in predictive modeling and has become an important framework in both statistics and machine learning. Given a random sample \( \mathcal{D}_n = \{(X_i, Y_i) \in \mathcal{X} \times \mathcal{Y}\}_{i=1}^n \) with exchangeable elements distributed as $(X,Y)$, and a regression function \( m: \mathcal{X} \to \mathcal{Y} \), conformal inference provides a general framework to construct an estimated prediction region \( \widehat{C}^\alpha(\cdot) \) such that \( \mathbb{P}(Y \in \widehat{C}^\alpha(X)) \geq 1 - \alpha \), where $\mathbb{P}$  is the probability law over the random sample $\mathcal{D}_{n}$ and $(X,Y).$ This method offers non-asymptotic guarantees \cite{shafer2008tutorial,barber2023conformal}.
 However, the development of conformal inference for settings with non-standard data structures, such as functional data and graphs, remains limited \cite{lugosi2024uncertainty}.
 This is primarily due to challenges in defining a valid method in spaces where the notion of order is absent. This limitation can be addressed, for example, by incorporating depth measures into conformal inference algorithms.

To bridge this gap, we propose a novel framework that leverages the strengths of depth measures to define rankings in general spaces, combined with conformal inference, to create prediction regions using robust algorithms for general regression models with non-asymptotic guarantees. From an asymptotic perspective, our methods offer fast non-parametric rate, due to the incorporation of kernel mean embedding as a depth measure   \cite{optimalrates}.
 Technically, our approach extends to multivariate or separable Hilbert space models within the distributional conformal framework proposed by \cite{chernozhukov2021exact}.
\subsection{Motivating application: Personalization of  physical activity}
\begin{figure}[ht!]
    \centering
    
    \begin{subfigure}[b]{0.5\textwidth}
        \includegraphics[width=\textwidth,height=0.65\textwidth]{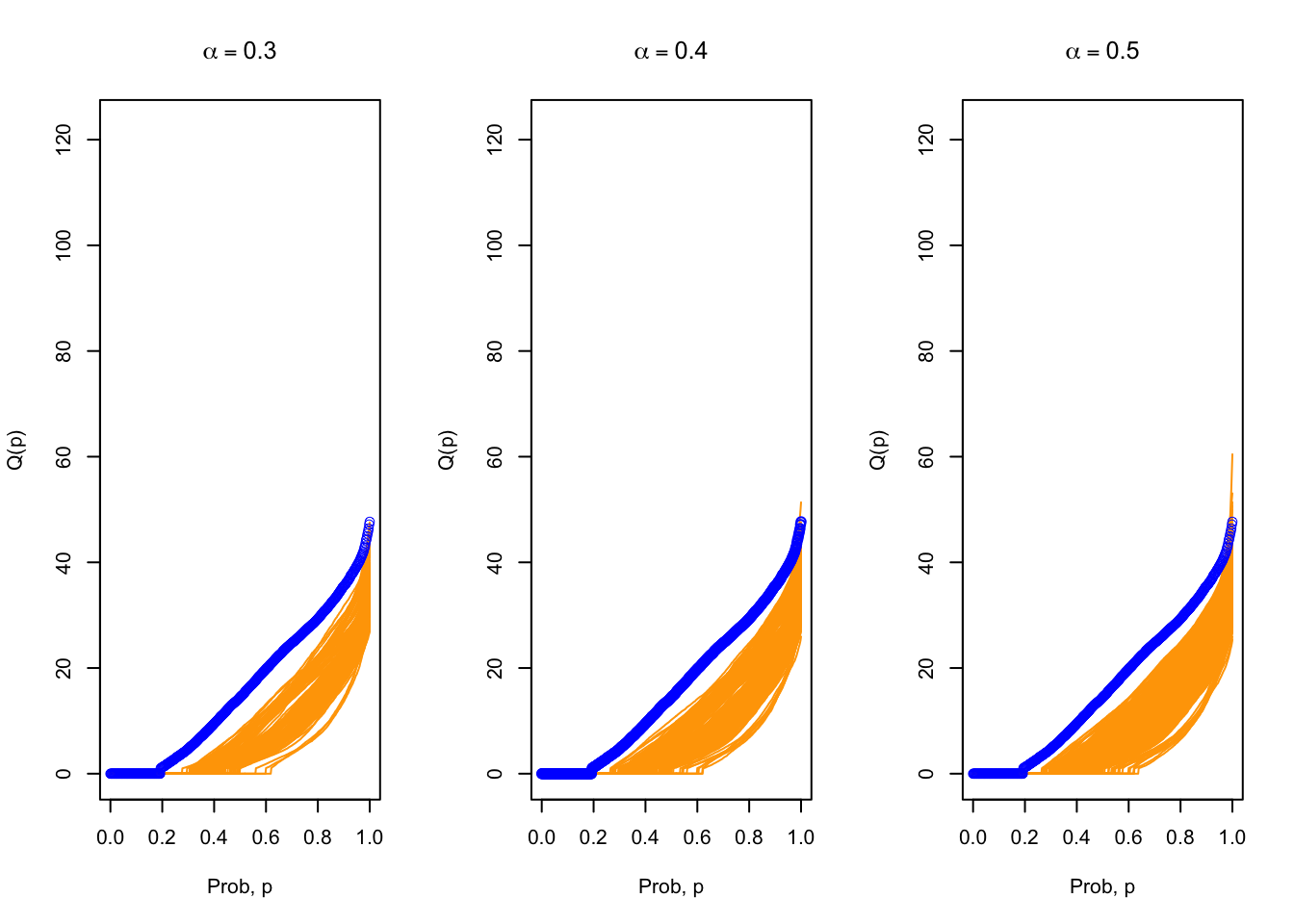}
        \caption{More active than expected}
        \label{fig:figura1}
    \end{subfigure}
    
    \begin{subfigure}[b]{0.5\textwidth}
        \includegraphics[width=\textwidth,height=0.65\textwidth]{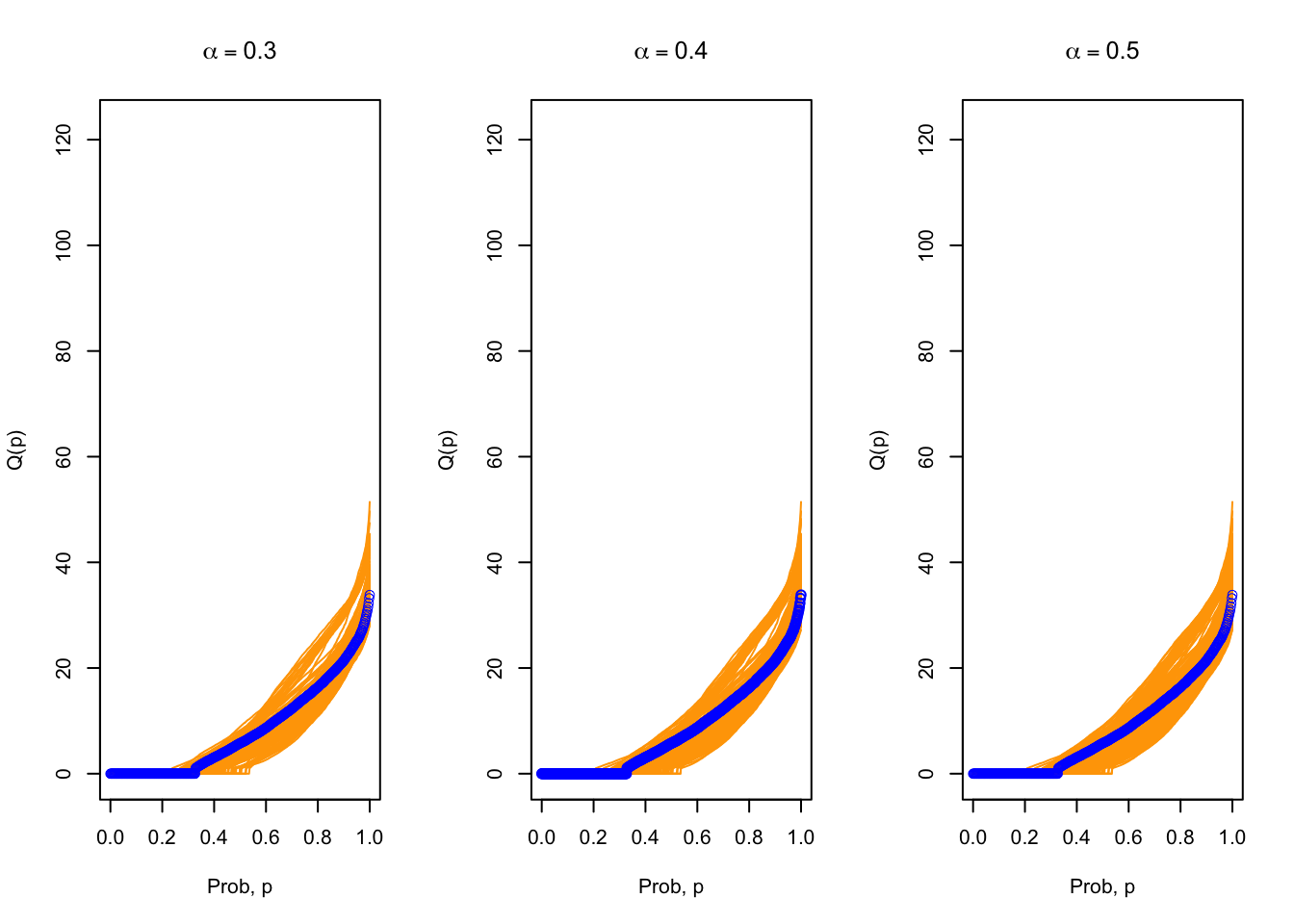}
        \caption{Active in the range}
        \label{fig:figura2}
    \end{subfigure}
    
    \begin{subfigure}[b]{0.5\textwidth}
        \includegraphics[width=\textwidth,height=0.65\textwidth]{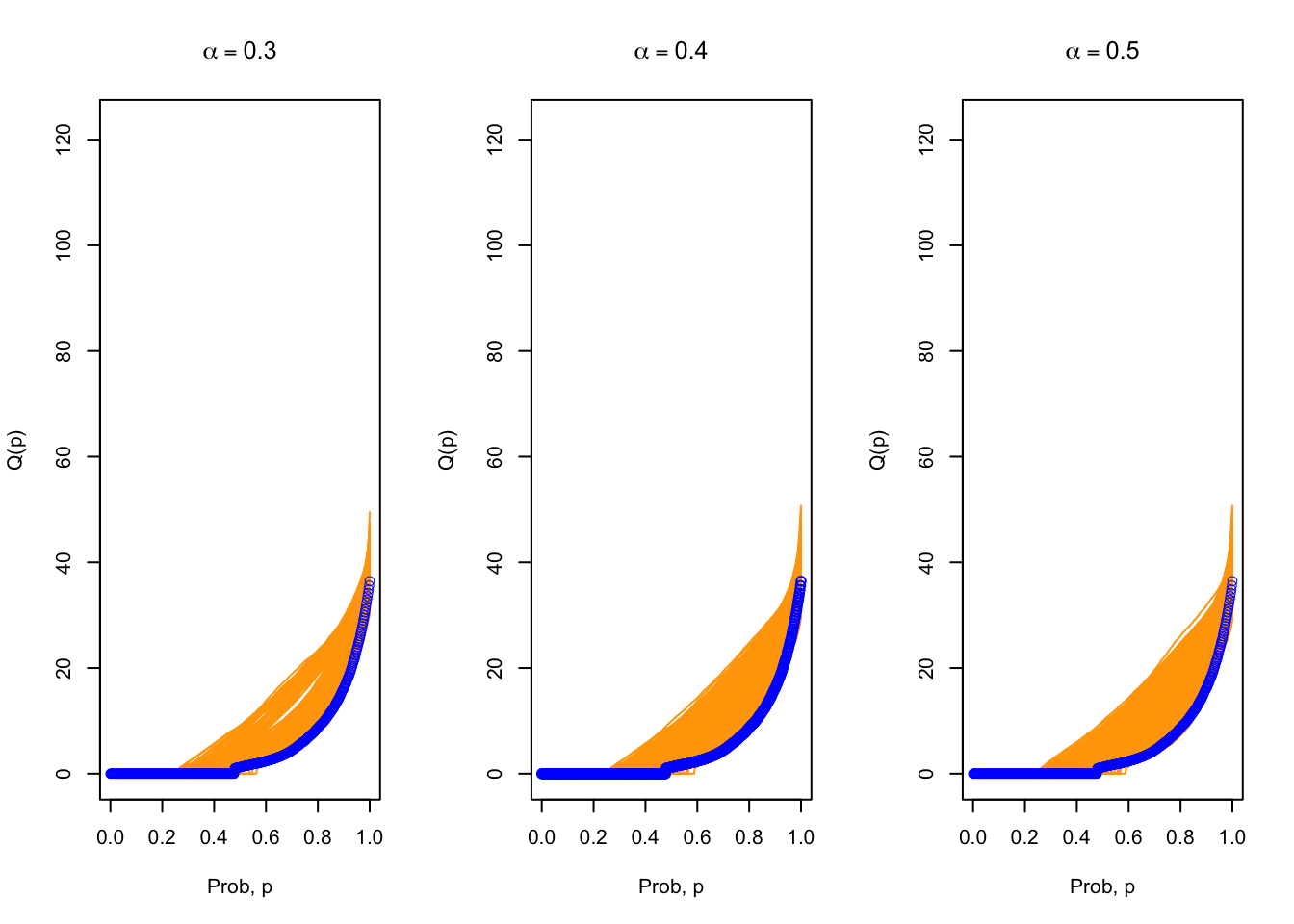}
        \caption{Less active than expected}
        \label{fig:figura3}
    \end{subfigure}
    
    \caption{Physical activity behavior of three approximately 50 years females  (a, b, c)  old with different body mass index and age. Blue lines represent  real physical activity, and shaded areas denote the prediction regions at different confidence levels $\alpha$.}
    \label{fig:tres}
\end{figure}
 The digital and precision medicine revolution has sparked a transformative era in healthcare, bringing forth advanced and personalized clinical interventions  \cite{kosorok2019precision,  matabuena2024multilevel}.  This paradigm shift empowers healthcare providers to harness the potential of modern wearable devices for continuous monitoring of specific patient characteristics, opening avenues for more sophisticated phenotyping and tailored treatments. One domain that particularly stands out to greatly benefit from these technological advancements is physical activity \cite{matabuena2022physical}.

From a clinical perspective, physical activity represents a low-cost intervention \cite{matabuena2023distributional},
 with great potential on the global healthcare system across diseases. The integration of wearable devices allows for comprehensive tracking of patients' physical activity at high resolution-level, enabling healthcare professionals to gain insights into their overall health and wellness.

This study is motivated by the need for more personalized recommendations of physical exercise based on individual characteristics using the novel prediction regions algorithm. Although there is a growing movement towards personalized physical activity prescriptions in the field of physical activity, current medical guidelines often provide standardized recommendations, such as weekly aerobic exercise ranging from 150 to 300 minutes \cite{piercy2018physical}. In order to illustrate the intuition behind our method, Figure \ref{fig:tres} displays the observed physical quantile trajectories of physical activity for three females individuals in blue, along with the estimated prediction regions at various confidence levels $\alpha \in (0,1)$. These regions are calculated using healthy individuals from the US. population as a reference. We observe that individual b falls within the expected range; however, individuals a and c are more and less active, respectively, than expected. Creating a predictive region conditioned to the patients characteristics will provide a formal method to actualized individualized physical activity.

\subsection{Notation and problem definition}

Consider a random sample $\mathcal{D}_{n}=\{(X_1,Y_1),\dots,(X_n,Y_n)\}$ of $n$ independent copies of the random variables $(X,Y)\sim P$, where $X$ takes values in the space $\mathcal{X}$ and $Y$ takes values in the space $\mathcal{Y}$. Here, we assume that $\mathcal{X}$ and $\mathcal{Y}$ are separable Hilbert spaces. Let $k_{\mathcal{X}}: \mathcal{X} \times \mathcal{X} \rightarrow \mathbb{R}$ be a positive definite kernel on $\mathcal{X}$, inducing a reproducing kernel Hilbert space (RKHS) $\mathcal{H}_{\mathcal{X}}$ with an inner product $\langle \cdot, \cdot \rangle_{\mathcal{X}}$. Similarly, let $k_{\mathcal{Y}}: \mathcal{Y} \times \mathcal{Y} \rightarrow \mathbb{R}$ be defined, and denote the related RKHS as $\mathcal{H}_{\mathcal{Y}}$. To avoid technical problems, we assume that $k_{\mathcal{X}}$ and $k_{\mathcal{Y}}$ are universal kernels on $\mathcal{X}$ and $\mathcal{Y}$, respectively, and then, $\mathcal{H}_{\mathcal{X}}$, $\mathcal{H}_{\mathcal{Y}}$ are simultaneously dense on $\mathcal{X}$ and $\mathcal{Y}$, see \cite{sriperumbudur2011universality}.

In this paper, for simplicity, we restrict our attention to a regression model defined as follows:

\begin{equation}\label{eqn:model}
    Y= m(X)+\sigma(X)\epsilon= m(X)+\epsilon(X),
\end{equation}

\noindent where $\epsilon(\cdot) \in \text{GP}(0_{\mathcal{H}_{\mathcal{Y}}},\Sigma)$ is the random error specified by a Gaussian process, $m(\cdot)\in  \mathcal{H}_{\mathcal{Y}}$ is the regression function, and $\sigma(X)\in \mathcal{H}_{\mathcal{Y}}$  introduces the heteroscedastic nature in the random error. In the case that $\sigma(X)= \sigma,$  we say the regression model defined in Equation \ref{eqn:model} is homoscedastic, and we denote $\epsilon(X)$ simply as $\epsilon$. For identifiability purposes, we suppose that  $\mathbb{E}(\epsilon(X)|X)=0$ and $\text{Cov}(\epsilon(X),\epsilon(X))=\Sigma(X)$.

In our specific context, we define the population prediction region of probability level $\alpha \in (0,1)$ as a subset of the joint space $\mathcal{X} \times \mathcal{Y}$ that holds the true response $Y$ for a given predictor $X$ with probability $1-\alpha$, that is $P(Y\in C^{\alpha}(X)|X)=  1 - \alpha$ for all $(X,Y) \in \mathcal{X} \times \mathcal{Y}$. For simplicity, given a random pair $(X,Y)$, we assume that such a predictive region exists. For a practical purposes, we establish the convergence to certain  oracle prediction regions that we also denote by $C^{\alpha}(X)$ (that we define later).

For practical purposes in terms of computational feasibility and theoretical analysis, to estimate the predictive region $C^{\alpha}(x)$, we partition the observed random sample $\mathcal{D}_{n}$ into three disjoint subsets: $\mathcal{D}_{n}=\mathcal{D}_{\text{train}} \cup \mathcal{D}_{\text{train}{2}} \cup \mathcal{D}_{\text{test}}$. The index sets are denoted as $\mathcal{S}_{1}:= \{ i\in [n]: (X_i, Y_i)\in \mathcal{D}_{\text{train}} \}$, $\mathcal{S}_{2}:= \{ i\in [n]: (X_i, Y_i)\in \mathcal{D}_{\text{train}{2}} \}$, $\mathcal{S}_{3}:= \{ i\in [n]: (X_i, Y_i)\in \mathcal{D}_{\text{test}} \}$, where $[n]:= \{1,\dots, n\}$.

Our primary goal is to propose a novel methodology for constructing estimators for prediction regions $C^{\alpha}(X)$, assuming the true underlying generative model specified for the scale-localization regression model specified in Equation (\ref{eqn:model}). To achieve this goal, we introduce a model-free approach based on statistical learning in RKHS capable of operating with more general generative models specified for higher--order moments. Consequently, our approach avoids the need to estimate the underlying function $m$ directly and instead incorporates information from the regression model in the construction of the estimator of prediction regions $C^{\alpha}(X)$ through a model-free approach.

	\subsection{Summary of contributions}
\begin{enumerate}
    \item We present a novel methodology for uncertainty quantification, leveraging the concept of conditional kernel mean embeddings     \cite{muandet2017kernel}. 

 Our algorithms operate when the response and predictors take in a separable Hilbert space, or event in a general metrics space of negative type \cite{schoenberg1937certain,schoenberg1938metric}. To illustrate this adaptability, we showcase an application in the realm of probability distribution of physical activity time series. Here, we employ the 2-Wasserstein distance as a representative metric, akin to working with quantile functions derived from  physical activity probability distributions.

     \item We propose conformalized versions \citep{vovk2005algorithmic} of novel uncertainty quantification algorithms, offering practitioners reliability through specific non-asymptotic guarantees of the type \(\mathbb{P}(Y \in \widehat{C}^{\alpha}(X)) \geq 1 - \alpha\), where $\mathbb{P}$  is the probability law over the random sample $\mathcal{D}_{n}$ and $(X,Y).$ Specifically, we introduce two algorithms tailored to the underlying signal-to-noise ratio of the regression function: homoscedastic and heteroscedastic settings.   In both cases, under some regularity conditions, for a fixed point \(x \in \mathcal{X}\), we provide asymptotic conditional consistency results of the form:

     \begin{equation}
    \mathbb{P}(Y \in \widehat{C}^{\alpha}(X) \mid X = x) = 1 - \alpha + o_{p}(1).
\end{equation}


    \item We introduce a novel bootstrap approach leveraging the asymptotic Gaussian structure of our estimators to define conditional tolerance predictive regions in probability from a random objects in separable Hilbert space.
    This extends the seminal work with functional data
    in the unconditional case by \cite{rathnayake2016tolerance}.
    \item We demonstrate the advantages of our novel models across various data structures, including multivariate Euclidean responses, distributional representations with the $2$-Wasserstein distance, standard functional-to-functional data analysis. To the best of our knowledge, our methodology uniquely fills a void in existing literature by offering a comprehensive framework for quantifying uncertainty within regression models of a functional nature, where no other general predictive framework currently exists.
    \item In the physical activity application:
    
    \begin{enumerate}
        \item We adapt our proposed algorithm
        to handle the survey design of the NHANES study.
        \item We provide a functional definition of recommended physical exercise levels over the full range of intensities recorded by the accelerometer device. An essential point in the modeling task is that definitions are given from a personalized standpoint, conditioned on factors such as sex, age, and BMI, which dramatically influence daily energy expenditure. This is a significant advancement over existing  guidelines that provide standart recommendations about physical activity without conditioning on specific patient characteristics.
    \end{enumerate}
    
\end{enumerate}

	\section{Uncertainty Quantification algorithms}

	\subsection{Tolerance regions and depth bands}

The study of tolerance regions has a long-standing tradition in statistics \cite{li2008multivariate}. For a random variable $Y \sim P$, defining tolerance regions typically involves determining a statistical region $T$ such that $\int_T P(dy) \geq \alpha$. Here, the region $T$ encompasses at least a fraction $\alpha$ (where $\alpha \in (0,1)$) of the probability mass of the random variable $Y.$ The final modeling goal is to define a tolerance region \( T \) such that the coverage satisfies the nominal value \( \alpha \in (0,1) \), while subject to geometric constraints, such as minimizing the hypervolume of \( T \), to optimize the region's shape and size.

Originally, tolerance regions were conceptualized under the assumption of Gaussian distributions \cite{murphy1948non,lucagbo2023rectangular}. As the field progressed, methodologies expanded to include semi-parametric \citep{liu2009building} and non-parametric approaches \citep{young2020nonparametric},
 primarily dealing with standard multivariate Euclidean data. However, with advancements in technology, especially in fields like medical science, there is now a demand for new methods for random objects such  as functions, graphs, or distributions that take values in a general metrics spaces \cite{lugosi2024uncertainty}.

Our research introduces new algorithms for random objects that take values in abstract spaces. Next, following \cite{li2008multivariate}, we introduce  two definitions of tolerance regions for the unconditional case.

\begin{definition} For  $ \alpha  \in (0,1)$,
 a random region $T(Y_1, \dots, Y_n)$ is a $\alpha$-content tolerance region (Type I) at confidence level $\gamma\in (0,1)$ if it satisfies:
  \begin{equation*}
    \mathbb{P}(P(Y \in T(Y_1, \dots, Y_n)) \geq \alpha \,|\, \mathcal{D}_n) = \gamma,
  \end{equation*}
\end{definition}

\noindent where $\mathbb{P}$ is the joint probability distribution of the random sample $\mathcal{D}_n$. 

\begin{definition}
  A region $T(Y_1, \dots, Y_n)$ is a $\alpha$-expectation tolerance region (Type II)  if it satisfies:
  \begin{equation*}
    \mathbb{E}(P(Y \in T(Y_1, \dots, Y_n)) \,|\, \mathcal{D}_n) = \alpha.
  \end{equation*}
\end{definition}

\begin{remark}
  In both definitions, $P$ denotes the probability distribution of the random variable $Y$, which takes values in an arbitrary separable Hilbert space $\mathcal{Y}$. The $\alpha$-parameter specifies the proportion of the population the tolerance region should cover, while the confidence level $\gamma$ reflects the level of uncertainty in the estimate.
\end{remark}
\subsubsection{Depth bands to define tolerance unconditional regions}

The cornerstone of our algorithm is the concept of data depth, a measure crucial for analyzing the structure of data in complex spaces. Consider a random variable \(Y\) in a space \(\mathcal{Y}\) with distribution \(P\). We define data depth as a function \(D(\cdot; P)\) that ranks each point \(y\) in the support of \(Y\) according to its proximity to \(Y\)'s geometric center. This center, denoted by \(\overline{y}\), is defined as \(\overline{y} = \arg\max_{y \in \mathcal{Y}} D(y; P)\). Points near \(\overline{y}\) are deemed to have higher depth, reflecting their centrality, whereas points farther from the center have depths approaching zero, indicating extreme or outlier points. 

Furthermore, we incorporate an axiomatic topological characterization of data depth, as proposed by \cite{nieto2016topologically}. This framework provides rigorous guidelines for defining and evaluating depth measures, ensuring that they accurately reflect the underlying data's topological and geometrical properties.

\begin{definition}
A statistical depth measure is a mapping $D: \mathcal{Y} \times \mathcal{P} \to [0,1]$, where $\mathcal{P}$ is a determined space of probability measures over $\mathcal{Y}$, that satisfies the following properties:
i) Distance invariance of $D$; ii) Maximality of $D$ at the center; iii)
Monotonicity of $D$ relative to the deepest point;
iv) Upper semi-continuity of $D$ in any random element $y \in \mathcal{Y}$; v) Receptivity of $D$ to the convex hull width across the domain; vi) Continuity of $D$ in $\mathcal{P}$.  

\end{definition}

In the Appendix $S1$, we provide a rigorous mathematical definition of the properties $i)$ to $vi)$.


Data depths, as proposed by  \cite{li2008multivariate}, provide a robust framework for establishing unconditional tolerance bands for multivariate data. These methods generalize the notion of order statistics and ranks for multivariate data.

 For simplicity, suppose that the probability law \(P\) is continuous. Consider an affine-invariant depth measure \(D(\cdot; P)\). Let \(D(Y_1; P), \dots, D(Y_n; P)\) represent the evaluations of \(n\) random observations \(\mathcal{D}_n = \{Y_i\}_{i=1}^{n}\), independently and identically distributed (i.i.d.) from \(P\). Define \(D^{(i)}\) for \(i = 0, \dots, n\), as the \(i\)th  reverse-order statistic from \(\{D(Y_i; P) \in [0,1]\}_{i=1}^{n}\), and set \(D^{(0)} = 1\), for convenience (see more details \cite{li2008multivariate}). Here, \(D^{(0)}\) represents the maximal depth value, akin to the geometric median, in its role as the most central point under the depth measure.


Based on this data sorting, we define different subspaces of \(\mathcal{Y}\) according to the following depth spacings algorithm:

\begin{itemize}
    \item \(MS_i = \{y : D^{(i-1)} > D(y; P) \geq D^{(i)}\}\), for \(i=1, \dots, n\), with \(D^{(0)}:= 1\),
    \item \(MS_{n+1} = \{y : D(y; P) < D^{(n)}\}\).
\end{itemize}

Next, we introduce formal results  from \cite{li2008multivariate} that relate the evaluations of depth-measures with their probability coverage.

\begin{theorem}\label{theorem:spacing}(\cite{li2008multivariate})
    Suppose \(\mathcal{D}_{n}= \{Y_1, \dots, Y_n\}\) is an i.i.d. sample from \(P\), and \(D(\cdot;P)\) is an affine-invariant depth measure. Then, 
    \begin{equation*}
        \left(  P
(MS_1),   P
(MS_2), \dots,   \mathbb{P}
(MS_{n})\right) \sim  \left(  
Z_{(1)},  
Z_{(2)}-Z_{(1)}, \dots,   
Z_{(n)}-Z_{(n-1)}\right),
    \end{equation*}
  \noindent  where the random sample \(\mathcal{S}_{n}=\{Z_{1},\dots, Z_{n}\}\) is an i.i.d.
  random elements distributed as \(U[0,1]\), and \(Z_{(i)}\) is the \(i\)th order element from \(\mathcal{S}_{n}\).
\end{theorem}

For each \(r_{n} \in \{0, 1, \dots, n\}\), the properties of tolerance regions are influenced by the characteristics of the sequence of univariate random variables \(\mathcal{S}_{n}\).
 This paper proposes defining the tolerance region \(\widehat{T}^{r_{n}}(Y_{1},\dots, Y_{n})\) as the union of a selected number of inner spacings, given by:

\begin{equation}
    \widehat{T}^{r_{n}}(Y_{1},\dots, Y_{n}) = \bigcup_{i=1}^{r_n} MS_i =\bigcup_{i=1}^{r_n} \{ y : D(y; P) \geq D^{(i)} \},
\end{equation}
\noindent where \(MS_i\) represents the \(i\)-th inner spacing, defined by the depth measure \(D(y; P)\)  with respect to the distribution \(P\). Drawing on Theorem \ref{theorem:spacing},  we can infer the distinct distributional properties of the tolerance regions \(\widehat{T}^{r_{n}}(Y_{1},\dots, Y_{n})\) for each \(r_n=1,\dots,n\).

\begin{corollary} (\cite{li2008multivariate})
Under the conditions of Theorem \ref{theorem:spacing}, we have

\begin{equation*} 
  P(Y\in \widehat{T}^{r_{n}}(Y_{1},\dots, Y_{n})\mid \mathcal{D}_{n})
  \sim Beta(r_n, n+1-r_n), \quad for \; n=1,\dots,n.
\end{equation*}
\end{corollary}

 \begin{remark}
 In the case of $\alpha-$tolerance regions in expectation $\mathbb{E}(\text{Beta}(n-2r+1, 2r)) = \alpha$,  we have an analytical solution given by $r = (n+1)\alpha$.
 \end{remark}

\subsubsection{Kernel mean embedding and integrated depth bands}

 This
 paper exploits the concept of depth bands, particularly focusing on the $h$-integrated depth band, which, mathematically, corresponds to the notion of kernel mean embedding. Kernel mean embeddings are based on the naive idea 
to map distributions into a reproducing kernel Hilbert space (RKHS)
in which the whole arsenal of kernel methods can be extended to probability measures. The kernel mean embeddings
can be viewed as a generalization of the original
“feature map” common to support vector machines (SVMs) and other
kernel methods.

For any positive definite kernel function $k_{\mathcal{Y}}: \mathcal{Y} \times \mathcal{Y} \rightarrow \mathbb{R}$, there exists a unique reproducing kernel Hilbert space (RKHS) $\mathcal{H}_{\mathcal{Y}}$. RKHS is a space, often infinite-dimensional, of functions $h: \mathcal{Y} \rightarrow \mathbb{R}$ where evaluation can be expressed as an inner product. Particularly,  $h(f)=\langle h, k_{\mathcal{Y}}(\cdot, f)\rangle_{\mathcal{H}_{\mathcal{Y}}}$ for all $h,f \in \mathcal{H}_{\mathcal{Y}}$.

For a given marginal distribution $P$, the kernel mean embedding (KME) $\mu_{P}$ is defined as the expectation of the feature $k_{\mathcal{Y}}(\cdot,Y)$:
$$
\mu_{P}=\mathbb{E}_{Y}[k_{\mathcal{Y}}(\cdot,Y)] \in \mathcal{H}_{\mathcal{Y}}
$$

\noindent and always exists for bounded kernels. Furthermore, for characteristic kernels, these embeddings are injective, uniquely defining the probability distribution (\cite{sriperumbudur2008injective,sriperumbudur2010hilbert}). Popular kernels like Gaussian and Laplace kernels possess this property.

The second-order mean embeddings, also termed covariance operators in \cite{fukumizu2004dimensionality}, are defined as the expectation of tensor products between features:
$$
C_{X Y}=\mathbb{E}_{X Y}[ k_{\mathcal{X}}(\cdot, X) \otimes  k_{\mathcal{Y}}(\cdot, Y)],
$$ 

\noindent where $\otimes$ represents the tensor product, and they always exist for bounded kernels. Covariance operators generalize the familiar notion of covariance matrices to accommodate infinite-dimensional feature spaces.

For a set of i.i.d. samples $Y_{1}, \ldots, Y_{n}$, the kernel mean embedding is typically estimated by its empirical version

\begin{equation*}
\widehat{\mu}_{P}=\frac{1}{n} \sum_{i=1}^{n} k\left(\cdot, Y_{i}\right) \tag{2}
\end{equation*}

\noindent from which various associated quantities, including the estimators of the squared RKHS distances between embeddings needed for kernel-based hypothesis tests, can be derived.

From a mathematical statistics perspective, kernel mean embeddings are considered a special case of 
h-depths \cite{SCOVEL2010641,wynne2021statistical}.
 Explicit mathematical connections and relationships
 based on \cite{SCOVEL2010641} are introduced in the Supplemental Material $S_1$ (see complete description in Theorem 3 from \cite{wynne2021statistical}). 

Given the mentioned connection, an empirical estimator of kernel mean embedding  can be considered in practice  as a data-depth measure. 

\begin{equation*}
\widehat{D}_{k}(\cdot;P) = \frac{1}{n}\sum_{i=1}^{n} k(\cdot,Y_i).
\end{equation*}

Then, the Algorithm\ref{alg:alpha_expectation_tolerance}
  can be utilized to obtain $\alpha$-expectation tolerance regions $\widehat{T}(Y_1,\dots,Y_n)$.

\begin{algorithm}[H]
  \SetAlgoLined
  \KwIn{Dataset $\{Y_j\}_{j=1}^{n}$}
  \KwOut{$\alpha$-expectation tolerance region $\widehat{T}(Y_1,\dots,Y_n)$}
  
  \begin{enumerate}
    \item Estimate the kernel-based depth for each observation $Y_j$ in the dataset $\{Y_j\}_{j=1}^{n}$:
    \[
    \widehat{D}_{k}(Y_j;P) = \frac{1}{n} \sum_{i=1}^{n} k(Y_j,Y_i),
    \]
    where $k(\cdot,\cdot)$ denotes the kernel function.
    \item Sort the estimated depths $\{\widehat{D}_{k}(Y_j;P)\}_{j=1}^{n}$ in non-increasing order to obtain the order statistics $\widehat{D}^{(1)} \geq \widehat{D}^{(2)} \geq \cdots \geq \widehat{D}^{(n)}$.
    \item Define the $\alpha$-expectation tolerance region $\widehat{T}(Y_1,\dots,Y_n)$ as the set of points $Y$ satisfying:
    \[
    \widehat{T}(Y_1,\dots,Y_n) = \{Y: \widehat{D}_{k}(Y;P) \geq \widehat{D}^{\left\lceil n+1-\alpha(n+1) \right\rceil}\},
    \]
   \noindent where $\left\lceil \cdot \right\rceil$ denotes the ceiling function, ensuring that the index is an integer.
  \end{enumerate}
  \caption{Estimation of $\alpha$-Expectation Tolerance Regions}
  \label{alg:alpha_expectation_tolerance}
\end{algorithm}

\subsection{Kernel Conditional Mean Embeddings}

	In this section, we focus on the  notion of kernel conditional mean embedding.  We assume that we have access to a random sample $\mathcal{D}_n = \{(X_i, Y_i)\}_{i=1}^{n}$ from the random variable $(X, Y) \sim P$. Now, we define the concept of conditional $\alpha$-expectation tolerance region.

\begin{definition}
For  $\alpha \in (0,1)$, and fixed $X=x$, the region $T_{x}((X_1, Y_1), \dots, (X_n, Y_n))$ is called a local $\alpha$-expectation tolerance region (or a prediction region) if

\begin{equation*}
    \mathbb{E}(P(Y \in T_{x}((X_1, Y_1), \dots, (X_n, Y_n)) | \mathcal{D}_n, X=x) = \alpha,
\end{equation*}
\noindent where $\mathbb{E}$ is the expectation with respect to the random sample $\mathcal{D}_n = \{(X_i, Y_i)\}_{i=1}^{n}$.
\end{definition}

Similarly to the previous section, a natural way to define conditional tolerance regions is by means of kernel mean embedding, but,  via their conditional version.

Roughly speaking, Conditional Kernel Mean Embeddings (CKMEs) extend the concept of kernel mean embedding to capture the conditional distribution of the pair $(X, Y)$ based on the idea of embedding the distribution in a specific Reproducing Kernel Hilbert Space (RKHS) \cite{song2009hilbert, muandet2017kernel}. In other words, CKMEs provide a way to represent the relationship between input and output variables in a probabilistic framework, allowing for tasks such as conditional density estimation, conditional distribution regression, and conditional hypothesis testing.

The CKME of a conditional distribution $P(Y|X=x)$ is a function $\mu_{P(Y|X=x)} \in \mathcal{H}_{\mathcal{Y}}$ such that for any bounded and continuous function $f:\mathcal{Y} \rightarrow \mathbb{R}$, the conditional expectation can be computed as:

\begin{equation}
    \mathbb{E}[f(Y)|X=x] = \langle f, \mu_{P(Y|X=x)} \rangle_{\mathcal{H}_{\mathcal{Y}}},
\end{equation}

where $\langle \cdot, \cdot \rangle_{\mathcal{H}_{\mathcal{Y}}}$ is the inner product in the RKHS $\mathcal{H}_{\mathcal{Y}}$. The CKME thus provides a way to embed the conditional distribution in the RKHS.

With the aforementioned building blocks in place, we now extend KMEs to to the case of conditional distributions, defining kernel conditional mean embeddings (CMEs) \citep{song2009hilbert} as follows:

$$
\mu_{P(Y \mid X)}(x)=\mathbb{E}_{Y \mid x}[\phi(Y) \mid X=x] \in \mathcal{H}_ \mathcal{Y}
$$

requiring an operator $C_{Y \mid X}: \mathcal{H}_{\mathcal{X}} \rightarrow \mathcal{H}_{\mathcal{Y}}$ that satisfies: (a)  $\mu_{P(Y \mid X)}(x)=C_{Y \mid X} \psi(x)$, where $\psi(x)= {k}_{\mathcal{X}}(\cdot,x)$ and (b) $\left\langle g, \mu_{P(Y \mid X)}(x)\right\rangle_{\mathcal{H}_{\mathcal{Y}}}=\mathbb{E}_{Y \mid x}[g(Y) \mid X=x]$ for $g \in \mathcal{H}_{\mathcal{Y}}$. Assuming $\mathbb{E}_{Y \mid x}[g(Y) \mid X=\cdot] \in \mathcal{H}_{\mathcal{X}}$ \cite{shimizu2024neural}
 the following operator satisfies the requirements:

$$
C_{Y \mid X}=\left(C_{X X}\right)^{-1} C_{X Y}.
$$

Given i.i.d. samples $\left\{\left(X_{i}, Y_{i}\right)\right\}_{i=1}^{n} \sim P$, an empirical estimate of the conditional covariance operator \cite{fukumizu2004dimensionality,https://doi.org/10.1002/cjs.11329}

$$
\begin{aligned}
\hat{C}_{Y \mid X} & =\left(\hat{C}_{X X}+\lambda I\right)^{-1} \hat{C}_{X Y} \\
& =\Phi\left(K_{X}+\lambda I\right)^{-1} \Psi^{\top}
\end{aligned}
$$

where $\lambda>0$ is a regularization parameter, $I$ is the $n\times n$ identity matrix,  $K_{X}$ is the Gram matrix $\left(K_{X}\right)_{i j}=k\left(x_{i}, x_{j}\right)$, and $\Psi$ and $\Phi$ are the feature matrices stacked by columns: $\Psi=\left[\psi\left(x_{1}\right), \ldots, \psi\left(x_{n}\right)\right]$ and $\Phi=\left[\phi\left(y_{1}\right), \ldots, \phi\left(y_{n}\right)\right]$.  Alternatively, this empirical estimate can be obtained by solving following function-valued regression problem \cite{grunewalder2012conditional}:

\begin{equation*}
\underset{C: \mathcal{H}_{\mathcal{X}} \rightarrow \mathcal{H}_{\mathcal{Y}}}{\arg \min } \frac{1}{n} \sum_{i=1}^{n}\left\|\phi\left(y_{i}\right)-C \psi\left(x_{i}\right)\right\|_{\mathcal{H}_{\mathcal{Y}}}^{2}+\lambda\|C\|_{H S}^{2}, \tag{1}
\end{equation*}

where $\|C\|_{H S}$ is the Hilbert-Schmidt norm. Putting together these elements, we arrive at the empirical estimator:

\begin{equation*}
\hat{\mu}_{P(Y \mid X)}(x)=\hat{C} \psi(x)=\sum_{i=1}^{n} \beta_{i}(x) \phi\left(y_{i}\right)=\Phi \boldsymbol{\beta}(x) \tag{2}
\end{equation*}

where:

$$
\boldsymbol{\beta}(x)=\left(K_{X}+\lambda I\right)^{-1} \boldsymbol{k}_{X}
$$

with $\boldsymbol{k}_{X}=\left[k\left(x_{1}, x\right), \ldots, k(\left(x_{n}, x\right)\right]^{\top}$. In contrast to KMEs for marginal distributions, CMEs employ nonuniform weights $\beta_{i}(x)$, which are not constrained to be positive or sum up to one.

 Using $\widehat{D}_{k}(\cdot;P_{Y|X=x})$, we define the contour levels by  $\widehat{C}^{\alpha}(x) = \{y \in \mathcal{Y} : \widehat{P}(\widehat{D}_{k}(y;P_{Y|X=x})\mid X=x) \geq \widehat{q}_{1-\alpha} \}$, where $\widehat{q}_{1-\alpha}$ is a quantile of calibration.

  

\subsection{Conditional Conformal depth bands algorithm with non-asymptotic guarantees}

In this section, we focus on using the theory of tolerance regions to propose an explicit prediction region algorithm, which incorporates conditional kernel mean embedding. We replace the notation $\alpha$-tolerance region $T(Y_1, \dots, Y_n)$ with $\widehat{\mathcal{C}}^\alpha(\cdot)$. Additionally, we define the population prediction region as  
\[ C^\alpha(x) = \{ y \in \mathcal{Y} : P(D_k(y; P_{Y\mid X=x}) \geq q_{1-\alpha}(x) \mid X=x) \}, \]
where $q_{1-\alpha}(x)$ is the $1-\alpha$ quantile of the probability distribution $g(x,y) = P(D_k(Y; P_{Y\mid X} \leq y \mid X=x)$.  In practice, under the continuity hypothesis, due to the properties of depth-measure and integral transformation rank, we have \( q_{1-\alpha}(x) = 1 - \alpha \).

\subsubsection{General formulation of our algorithm}

\begin{algorithm}[H]
  \SetAlgoLined
  \KwIn{Dataset $\mathcal{D}_n = \{(X_i, Y_i)\}_{i\in[n]}$, confidence level $\alpha \in (0,1)$}
  \KwOut{Prediction region $\widehat{C}^{\alpha}(x)=\{y \in \mathcal{Y}: \widehat{g}(x,y) \geq \widehat{q}_{1-\alpha}\}$}
  
  \begin{enumerate}
    \item Split $\mathcal{D}_n$ into three disjoint sets $\mathcal{D}_{train}$, $\mathcal{D}_{test}$, and $\mathcal{D}_{test2}$ \;
    \item Utilize the random sample $\mathcal{D}_{train}$ to estimate $\widehat{D}_k(\cdot;P_{Y|X})$\;
    \item Calculate $r_i = \widehat{D}_k(Y_i;P_{Y|X_i})$ for all $(X_i,Y_i) \in \mathcal{D}_{test}$\;
    \item Estimate the conditional probability distribution function of the random sample $\{(X_i,r_i)\}_{i\in[S_2]}$. Denote this estimator as $\widehat{g}(x,y)= \widehat{P}(\widehat{D}_k(Y;P_{Y|X}) \leq y \mid X=x)$;
    \item Calculate $s_i= g(X_i,Y_i)$ for all $(X_i,Y_i) \in \mathcal{D}_{test2}$\;
    \item Calculate the empirical quantile $1-\alpha$ with the random sample $\{s_{i}\}_{i\in \mathcal{D}_{test2}}$, denoted as $\widehat{q}_{1-\alpha}$\;
    \item Return the prediction region $\widehat{C}^{\alpha}(x)=\{y \in \mathcal{Y}: \widehat{g}(x,y) \geq \widehat{q}_{1-\alpha}\}$\;
  \end{enumerate}
  \caption{General prediction region algorithm}
      \label{algorithm:3sal}
\end{algorithm}

Algorithm \ref{algorithm:3sal} contains the core steps of our proposal. In the first split $\mathcal{D}_{train},$ we estimate the conditional depth measure. In the second split, $\mathcal{D}_{test}$, we estimate the conditional distributional of the conditional depth measure denote such estimators as  $\widehat{g}(x,y)= \widehat{P( \widehat{D}_k(Y;P_{Y|X})\leq y|X=x)}$. Finally, in the last split, we obtain the quantile of calibration $\widetilde{q}_{1-\alpha}$, and return  as a prediction region 
 $\widehat{C}^{\alpha}(x)=\{y \in \mathcal{Y}: \widehat{g}(x,y) \geq \widehat{q}_{1-\alpha}\}$. The data-split procedure guarantee non-asymptotic properties in the marginal coverage.

\begin{proposition}(Finite sample unconditional validity) Suppose that the random elements from $\mathcal{D}_{n}$ are iid (or exchangeable) and the conditional distribution estimator is invariant to permutations of the data independent the regression models is homoscedastic and heteroscedastic. Then    
\begin{equation}
    \mathbb{P}(Y\in \widehat{C}^{\alpha}(X))\geq 1-\alpha.
\end{equation}
\end{proposition}

\begin{remark}
The proof of Theorem is standard and a generalization of the results presented in \cite{vovk2005algorithmic} and \cite{kuchibhotla2020exchangeability} for three splits. 
\end{remark}

\begin{theorem} (Asymptotic unconditional validity) Suppose that $\mathcal{D}_{n}$ is a i.i.d sample, and the estimator of $g(x,y)$ is universal consistency uniform for all $y,x \in \mathcal{X}$, $\sup_{x\in \mathcal{X} } \sup_{y\in \mathcal{Y}}|g(x,y)-\widehat{g}(x,y)|= o_{p}(1)$. Then, as $n_{1},n_{2},n_{3}\to \infty$,
    \begin{equation}
    \mathbb{P}(Y\in \widehat{C}^{\alpha}(X))= 1-\alpha+o_{p}(1)
\end{equation}
\begin{proof}
See Appendix
\end{proof}
\end{theorem}
\begin{theorem} (Asymptotic conditional validity) Suppose that the estimator of $g(x,y)$ is uniform universal consistency  for all $y,x \in \mathcal{X}$, $\sup_{x\in \mathcal{X} } \sup_{y\in \mathcal{Y}}|g(x,y)-\widehat{g}(x,y)|= o_{p}(1)$. Then, for any $x\in \mathcal{X}$  as $n_{1},n_{2},n_{3}\to \infty$,  
    \begin{equation}
    \mathbb{P}(Y\in \widehat{C}^{\alpha}(X)\mid X=x)= 1-\alpha+o_{p}(1).
    \end{equation}
\end{theorem}
   
   \begin{proof}
See Appendix.
\end{proof}


\subsection{Homoscedastic case}

In the homoscedastic case, for any $x,x'\in \mathcal{X},$ $g(x,y)=g(x',y)$ $\forall y \in \mathcal{Y}$. Then, for simplicity, we denote $g(\cdot,y)$ to $g(y)$.   To adapt Algorithm \ref{algorithm:3sal}
to this scenario and increase the statistical efficiency, we can estimate the conditional distribution of the depth-band measure by utilizing the empirical distribution derived from the random elements $\{r_i\}_{i\in D_{test}}$. As a result, we no longer require three splits in the uncertainty quantification procedure.  Algorithm  \ref{algorithm:4} contain the core steps  of the proposed algorithm for such situation.

\begin{algorithm}[H]
  \SetAlgoLined
  \KwIn{Dataset $\mathcal{D}_n = \{(X_i, Y_i)\}_{i\in[n]}$, confidence level $\alpha \in (0,1)$}
  \KwOut{Prediction region $\widehat{C}^{\alpha}(x)=\{y \in \mathcal{Y}: \widehat{g}(x,y) \geq \widehat{q}_{1-\alpha}\}$}
  
  \begin{enumerate}
    \item Split $\mathcal{D}_n$ into three disjoint sets $\mathcal{D}_{train}$, $\mathcal{D}_{test}$\;
    \item Utilize the random sample $\mathcal{D}_{train}$ to estimate $\widehat{D}_k(\cdot;P_{Y|X})$\;
    \item Define $D(x,y)=  \widehat{D}_k(y;P_{Y|X=x})$\;
    \item Calculate $r_i = D(X_i,Y_i)$ for all $(X_i,Y_i) \in \mathcal{D}_{test}$\;
    \item Calculate the empirical quantile $1-\alpha$ with the random sample $\{r_{i}\}_{i\in \mathcal{D}_{test}}$, denoted as $\widehat{q}_{1-\alpha}$\;
    \item Return the prediction region $\widehat{C}^{\alpha}(x)=\{y \in \mathcal{Y}: D(x,y) \geq \widehat{q}_{1-\alpha}\}$\;
  \end{enumerate}
  \caption{Prediction region algorithm for the homocedastic case}
    \label{algorithm:4}
\end{algorithm}
\begin{assumption}
		Suppose that the following hold:
		\begin{enumerate}\label{1Smet}
			\item $n_1,n_2\to \infty.$
			\item $\hat{\mu}_{P(Y \mid X)}(x)$ is a consistent estimator in the sense that $\lim_{n_{1}\to \infty}
   \sup_{x\in \mathcal{X}} \norm{\hat{\mu}_{P(Y \mid X)}(x)-\mu_{P(Y \mid X)}(x)}_{\infty} \to 0$, in probability.
 \item  The population quantile \(q_{1-\alpha} = \inf\{t \in \mathbb{R} : G(t) = P(  {D}_k(Y;P_{Y|X})\geq  t) = 1-\alpha\}\) exists uniquely and is a continuity point of the function \(G(\cdot)\). In practice $q_{1-\alpha}=1-\alpha$.
 \end{enumerate}
	\end{assumption}


	\begin{theorem}\label{th:conshom}
		Under Assumption \ref{1Smet}, the estimated prediction region $\widehat{C}^{\alpha}(x;\mathcal{D}_{n})$ obtained using Algorithm \ref{algorithm:4} satisfies

		\begin{equation}
	\int_{\mathcal{X}} \mathbb{P}(Y\in \widehat{C}^{\alpha}(X)\triangle
 C^{\alpha}(X)\mid \mathcal{D}_{n},X=x) P_{X}(dx)= o_{p}(1), 
		\end{equation}

  \noindent where $P_{X}$ denote the probability law from $X.$
	\end{theorem}


\subsection{Heteroscedastic case}\label{sec:gamlss}
In the heteroscedastic regime, to obtain more parsimonious and interpretable models, we focus on the semi-parametric estimation of the conditional distribution $g(x,y)$. More specifically, we consider the GAMLSS family of models. Other non-parametric options are possible, such as the kNN conditional distributional algorithm (see, for example, \cite{dombry2023stone, matabuena2024knn}), but these are very sensitive to the dimension of the input space, as are other non-parametric estimators like the Nadaraya-Watson estimators.

\cite{rigby2005generalized}
 introduced the Generalized Additive Models for Location, Scale, and Shape (GAMLSS) framework, which extends traditional regression models to allow for the modeling  all parameters of a distribution, and not just the mean regression function. This approach provides a more flexible framework for statistical modeling, accommodating various types of data and distributions.

In GAMLSS models, the response variable, \(Y\), is specified by a parametric distribution, and the model defines relationships between the distribution parameters, for example,  (\(\boldsymbol{\mu}, \boldsymbol{\sigma}, \boldsymbol{\nu}, \boldsymbol{\tau}\)) and explanatory variables through monotonic link functions \(g_k\). These link functions relate the distribution parameters to linear predictors (\(\boldsymbol{\eta}_{k}\)), which are linear combinations of fixed effects (\(\mathbf{X}_{k} \boldsymbol{\beta}_{k}\)) and random effects (\(\sum_{j=1}^{J_{k}} \mathbf{Z}_{j k} \gamma_{j k}\)). The model is expressed as:

\[
g_{k}\left(\boldsymbol{\theta}_{k}\right)=\boldsymbol{\eta}_{k}=\mathbf{X}_{k} \boldsymbol{\beta}_{k}+\sum_{j=1}^{J_{k}} \mathbf{Z}_{j k} \gamma_{j k} \tag{1}
\]

\begin{itemize}
    \item \(\boldsymbol{\theta}_{k}\) are the parameters of interest (\(\boldsymbol{\mu}, \boldsymbol{\sigma}, \boldsymbol{\nu}, \boldsymbol{\tau}\)).    \ 
    \item  \(g_{k}\) are known monotonic link functions.     
\item \(\mathbf{X}_{k}\) and \(\mathbf{Z}_{j k}\) are known design matrices for fixed and random effects, respectively.
\item  \(\boldsymbol{\beta}_{k}\) and \(\gamma_{j k}\) are the coefficients for fixed and random effects, respectively.
\item \(\gamma_{j k}\) follows a multivariate normal distribution with mean \(\mathbf{0}\) and covariance matrix \(\mathbf{G}_{j k}^{-1}\).
\end{itemize}

This formulation allows for a highly flexible modeling approach, where each distribution parameter can be modeled using a set of explanatory variables, assuming a specified functional form. From a technical perspective, the theoretical analysis indicates that model parameter estimation relies on maximum likelihood equations. The related M-estimators preserve the properties of consistency and Gaussian's limit of the parameters, as well as parametric rates. Following \cite{van2007empirical},
we introduce some technical background about empirical process to establish the distributional convergence of $g(\cdot,\cdot)$ to a Gaussian process $Z$.
Let $Y_{1}, \ldots, Y_{n}$ be i.i.d. random elements in a measurable space $(\mathcal{Y}, \mathcal{A})$ with law $P$, and for a measurable function $f: \mathcal{Y} \rightarrow \mathbb{R}$ let the expectation, empirical measure and empirical process at $f$ be denoted by

$$
P f=\int f d P, \quad \mathbb{P}_{n} f=\frac{1}{n} \sum_{i=1}^{n} f\left(Y_{i}\right), \quad \mathbb{G}_{n} f=\sqrt{n}\left(\mathbb{P}_{n}-P\right) f
$$

Given a collection $\left\{f_{\theta, \eta}: \theta \in \Theta, \eta \in H\right\}$ of measurable functions $f_{\theta, \eta}: \mathcal{X} \rightarrow \mathbb{R}$ indexed by sets $\Theta$ and $H$ and "estimators" $\eta_{n}$, we wish to prove that, as $n \rightarrow \infty$,

\begin{equation*}
\sup _{\theta \in \Theta}\left|\mathbb{G}_{n}\left(f_{\theta, \eta_{n}}-f_{\theta, \eta_{0}}\right)\right| \rightarrow_{p} 0 \tag{1}
\end{equation*}

Here an "estimator" $\eta_{n}$ is a random element with values in $H$ defined on the same probability space as $Y_{1}, \ldots, Y_{n}$, and $\eta_{0} \in H$ is a fixed element, which is typically a limit in probability of the sequence $\eta_{n}$.

The result (1) is interesting for several applications. A direct application is to the estimation of the functional $\theta \mapsto P f_{\theta, \eta}$. If the parameter $\eta$ is unknown, we may replace it by an estimator $\eta_{n}$ and use the empirical estimator $\mathbb{P}_{n} f_{\theta, \eta_{n}}$. The result (11) helps to derive the limit behaviour of this estimator, as we can decompose

\begin{equation*}
\sqrt{n}\left(\mathbb{P}_{n} f_{\theta, \eta_{n}}-P f_{\theta, \eta_{0}}\right)=\mathbb{G}_{n}\left(f_{\theta, \eta_{n}}-f_{\theta, \eta_{0}}\right)+\mathbb{G}_{n} f_{\theta, \eta_{0}}+\sqrt{n} P\left(f_{\theta, \eta_{n}}-f_{\theta, \eta_{0}}\right) \tag{2}
\end{equation*}

If (1) holds, then the first term on the right converges to zero in probability. Under appropriate conditions on the functions $f_{\theta, \eta_{0}}$, the second term on the right

will converge to a Gaussian process by the (functional) central limit theorem. The behavior of the third term depends on the estimators $\eta_{n}$, and would typically follow from an application of the (functional) delta-method, applied to the map $\eta \mapsto\left(P f_{\theta, \eta}: \theta \in \Theta\right)$.

Assume that $\eta_{n}$ is "consistent for $\eta_{0}$" in the sense that

\begin{equation*}
\sup _{\theta \in \Theta} P\left(f_{\theta, \eta_{n}}-f_{\theta, \eta_{0}}\right)^{2} \rightarrow_{p} 0. \tag{3}
\end{equation*}
\begin{proposition}(\cite{van2007empirical})
Suppose that $H_{0}$ is a fixed subset of $H$ such that $\operatorname{Pr}\left(\eta_{n} \in H_{0}\right) \rightarrow 1$ and suppose that the class of functions $\left\{f_{\theta, \eta}: \theta \in \Theta, \eta \in H_{0}\right\}$ is $P$-Donsker. If (3) holds, then (1) is valid.
\end{proposition}

\begin{theorem} (Uniform consistency of GAMLSS model)
Suppose that $\mathcal{X}$ is a finite--dimensional space, the class of function related to kernel mean embedding is P-Donsker, and estimate the conditional distribution of the kernel mean embedding with a GAMLSS model. Then, the estimator $\widehat{g}(x,y)$ is uniform consistent for all $(x,y)\in \mathcal{X}\times \mathcal{Y},$
 
$ \sup_{x\in \mathcal{X} } \sup_{y\in \mathcal{Y}}|g(x,y)-\widehat{g}(x,y)|= o_{p}(1), \text{ and },  \sqrt{n}\left(\widehat{g}(\cdot,\cdot)-g(\cdot,\cdot)\right) \Rightarrow Z $ (in distribution)   
\noindent where $Z$ is a Gaussian process. 
\end{theorem}

\begin{remark}
There are very recent relevant researchers about the Donsker character of kernel mean embeddings, such as \cite{carcamo2024uniform} and \cite{pmlr-v201-park23a}. When the outcomes are defined in an infinite-dimensional space, the Donsker hypothesis can be violated. However, there are other arguments to justify the validity of bootstrap approaches in this environments (see, for example, \cite{martinez2024efficient}).

\end{remark}
\subsubsection{Tolerance regions in probability via bootstrapping} 
The literature on tolerance regions with functional data and complex statistical objects is scarce. For example, \cite{rathnayake2016tolerance} propose an unconditional tolerance region in probability using functional principal component analysis. However, the authors restrict their analysis to the supremum norm as the geometric selection and correct the numerical approximation with bootstrap to truncate and use estimates of eigenfunctions and eigenvalues, following prior work from \cite{goldsmith2013corrected}. Data from depths provide new opportunities to create more general algorithms and for a conditional and unconditonal perspective, for instance, using kernel mean embeddings beyond funtional data as case of graphs spaces. First, for a fixed $x\in \mathcal{X},$ we define the notion of a conditional tolerance region in probability.
   
 \begin{definition}
For a fixed $X=x,$  a random region $T_{x}(Y_1, \dots, Y_n)$ is a $\alpha$-content tolerance region (Type I) at confidence level $\gamma\in (0,1)$ if it satisfies:
  \begin{equation*}
    \mathbb{P}(P(Y \in T_{x}(Y_1, \dots, Y_n)) \geq \alpha \,|\, \mathcal{D}_n,X=x) = \gamma,
  \end{equation*}
\end{definition}

\noindent where $\mathbb{P}$ is the joint probability distribution of the random sample $\mathcal{D}_n$.

The basic idea of our algorithm is to exploit the asymptotic Gaussianity of the  conditional distributional $g(x,y)$ estimators indexed by a kernel mean embedding. Then, we combine this with Efron's and parametric (uniform) bootstrap to obtain tolerance regions of probability. The algorithm  \ref{algorithm:tol} contains the core steps of our proposal.

\begin{algorithm}[H]
  \SetAlgoLined
  \KwIn{Dataset $\mathcal{D}_n = \{(X_i, Y_i)\}_{i\in[n]}$, confidence level $\alpha \in (0,1)$, a point $x \in \mathcal{X}$.}
  \KwOut{Tolerance region in probability, $\widehat{C}^{\text{tolerance}, \alpha}(x) = \{y \in \mathcal{Y} : \widehat{g}(x, y) \geq \widehat{\widehat{q}}_{1-\alpha}\}$.}
  
  \begin{enumerate}
    \item Repeat $B$ times: Sample with replacement $n$ observations from $[n]=\{1, \dots, n\}$ to denote the set of indices $\mathcal{I}_b$, where $b=1, \dots, B$. Define $\mathcal{D}^b_n = \{(X_i, Y_i)\}_{i \in \mathcal{I}_b}$.
    \item Apply the bootstrap $B$ times with the different bootstrap databases $\mathcal{D}^b_n$ to estimate the prediction region at the confidence level $1-\alpha$ and obtain $B$ quantiles of calibration $\{\widehat{q}^b_{1-\alpha}\}_{b=1}^B$.
    \item Define $\widehat{F}_B^q(t) = \frac{1}{B} \sum_{i=1}^B 1\{\widehat{q}^b_{1-\alpha} \leq t\}$ and obtain the empirical quantile at $1-\alpha$, denoted $\widehat{\widehat{q}}_{1-\alpha}$.
    \item Return the tolerance region in probability $\widehat{C}^{\alpha, \text{tolerance}}(x) = \{y \in \mathcal{Y} : \widehat{g}(x, y) \geq \widehat{\widehat{q}}_{1-\alpha}\}$.
  \end{enumerate}
  \caption{Conditional Tolerance Region in Probability}
  \label{algorithm:tol}
\end{algorithm}




\section{Simulation study}
To demonstrate the versatility and desirable properties of our estimators, we concentrate on a simulation study encompassing different scenarios: i) multivariate Euclidean data; ii) functional to functional regression models; and iii) a probability distribution example. Given the arbitrary nature of selecting an estimator for the conditional distribution functions, our focus will be solely on the heteroscedastic context in the first scenario. 

\subsection{Multidimensional Euclidean data}
	
	\subsubsection{Homoscedastic case}
	
\noindent To evaluate the model's performance under homoscedastic conditions and controlled settings, we conducted a simulation study based on the classical multivariate Gaussian linear regression model. Specifically, we considered a $p$-dimensional Gaussian predictor variable $X = (X_1, \dots, X_p)^\top \sim \mathcal{N}((0, \dots, 0)^\top, \Sigma)$, where $(\Sigma)_{ij} = \rho \in [0,1]$ for all $i\neq j$ and $(\Sigma)_{ii} = 1$ for all $i=1,\dots,p$. Additionally, we considered an $m$-dimensional random response variable $Y = (Y_1, \dots, Y_m)^\top \in \mathbb{R}^m$, using the following functional relationship:

\begin{equation}
Y = m(X) + \epsilon = X^\top \beta + \epsilon,
\end{equation}

\noindent where $\beta$ is a $p \times m$ matrix of coefficients with all entries equal to 1. The random error $\epsilon = (\epsilon_1, \dots, \epsilon_m)^\top \sim \mathcal{N}(0, \mathrm{diag}(1, \dots, 1))$ is assumed to be independent from $X$ and satisfies $\mathbb{E}(\epsilon|X) = 0$.
\noindent To assess the performance of our model under various scenarios, we conducted a comprehensive simulation study with multiple cases. We explored different configurations for the values of $m$ and $p$, as well as various sample sizes, confidence levels, and correlation structures for the covariance matrix $\Sigma= \begin{pmatrix}
1 & \rho & \rho & \cdots & \rho \\
\rho & 1 & \rho & \cdots & \rho \\
\rho & \rho & 1 & \cdots & \rho \\
\vdots & \vdots & \vdots & \ddots & \vdots \\
\rho & \rho & \rho & \cdots & 1 \\
\end{pmatrix}$, governed by the parameter $\rho\in [0,1]$ . Specifically, we considered $m\in \{1,2,5\},$ $p\in \{2,5,10\}$, and the sample size $n \in \{500,2000,5000\}$. The confidence levels $\alpha$ ranged from $0.05$ to $0.5$ ($\alpha=\{0.05,0.1,0.2,0.5\}$), and we varied correlation parameter $\rho$ between $0, 0.5,$ and $0.75$.

\noindent To ensure the robustness of our findings, we employed a training set of size $|\mathcal{D}_{train}| = [n/2]$ and a test set of size $|\mathcal{D}_{test}| = n - |\mathcal{D}_{train}|$. Each scenario was repeated $500$ times, and we evaluated the finite marginal coverage in an additional test set $|\mathcal{D}_{test_{2}}| = 5000$ drawn from the same probability law.

\begin{figure}[ht!]
\centering
\includegraphics[width=1\linewidth , height=0.8\linewidth]{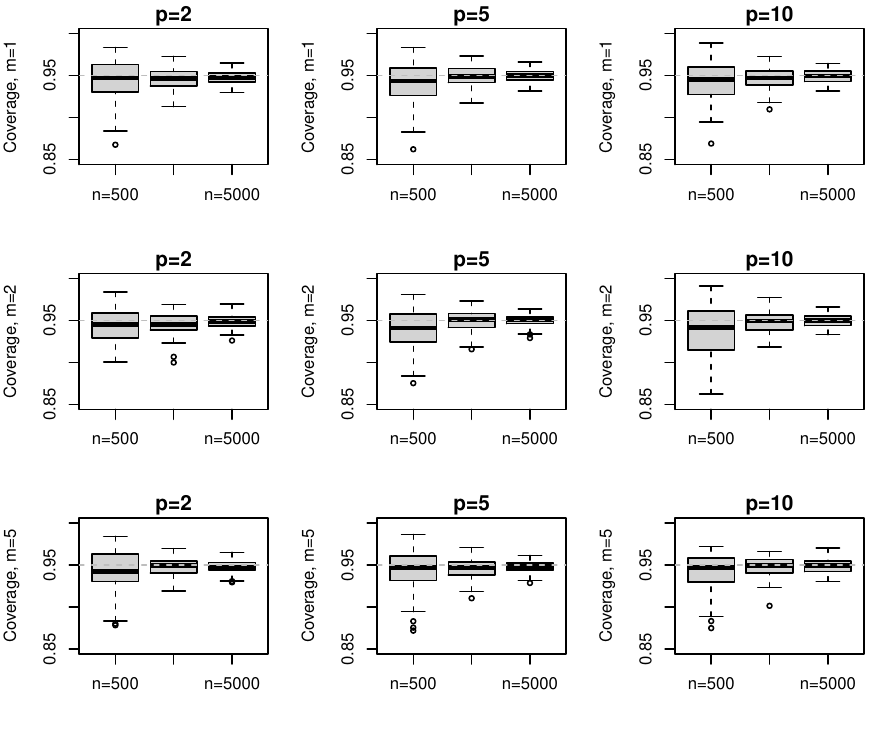}
\caption{Estimated marginal coverage, $\rho=0.5$, Euclidean data, homoscedastic case.}
\label{fig:figconf}
\end{figure}

Figure \ref{fig:figconf} shows the evaluation of marginal coverage $P(Y \in C^{\alpha}(X))$ in the test set. We demonstrate that the non-asymptotic properties are satisfied and, as the sample size increases, the variability of histogram is reduced, indicating the statistical consistency of the algorithm.

\begin{figure}[ht!]
\centering
\includegraphics[width=0.7\linewidth , height=0.5\linewidth]{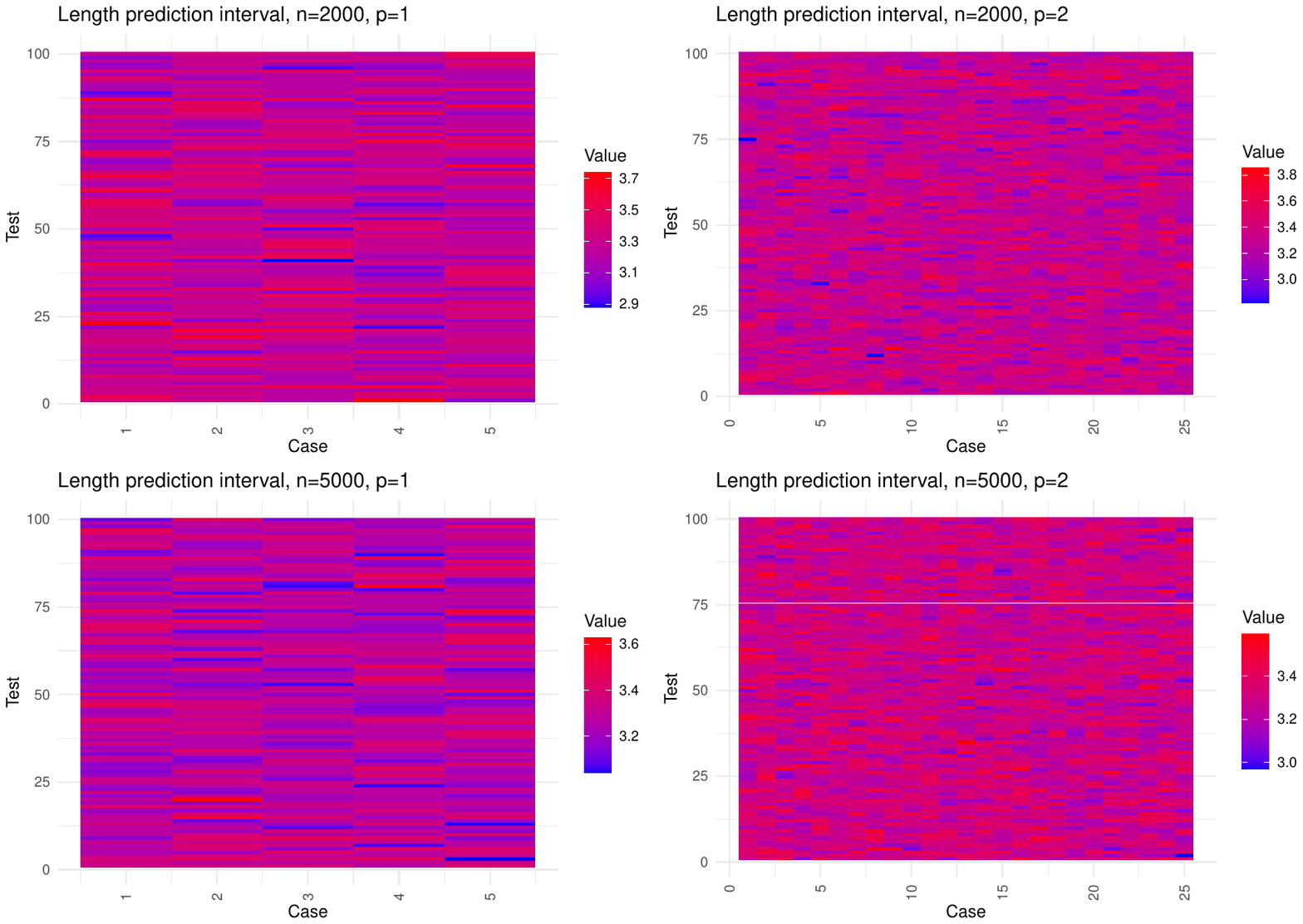}
\caption{Prediction interval length for $\alpha=0.9$ was calculated across 100 tests for the 5 and 25 scenarios (or $p=1,2$) combination of the 20th and 80th percentiles in the support of the predictor variable $X$. The results demonstrate consistent behavior across different combinations with a theoretical length of 3.28. With an increase in dimension, the variability also increases.}
 
\label{fig:stability}
\end{figure}

For the case where \(m=1\), our interest lies in comparing the stability of the prediction intervals obtained across different ranges of covariates, as well as comparing these to the length of the prediction interval, which, in this scenario, remains invariant regardless of the selected point. The analysis focuses on various percentiles for each coordinate of the covariate, considering the case \(p=1,2\). As we show in Figure  \ref{fig:stability}, our method closely approximates the theoretically optimal prediction interval length and demonstrates significant robustness against the selection of different configuration of covariation in the input space with a uniform pattern behavior.

	\subsubsection{Heteroscedastic case}
	
\noindent To account for heteroscedasticity in the simulation scenarios, we modify the previous simulation scheme and incorporate an additional term $\sigma: \mathbb{R}^{d}\to \mathbb{R}$ that represents the heteroscedastic term in the model. Specifically, we set the random error term to be $\sigma(X)\epsilon$, where $\sigma(x) = \norm{x}_2$ is a function that maps each feature vector to a scalar value corresponding to the $L^{2}$ norm. The overall generative model we consider is:

\begin{equation}
Y = m(X)+\sigma(X)\epsilon = X^T\beta + \norm{X}_2\epsilon .
\end{equation}

\noindent To predict the conditional mean function, we use linear regression without incorporating the heteroscedastic covariance structure in the model estimation. 

\noindent In this case, we use  the heterocedastic algorithm presented in  Algorithm \ref{algorithm:3sal},
 and utilize the GAMLSS   models describe in the Section \ref{sec:gamlss} with the beta distribution to estimate the conditional distribution, that assume the support of the response is $(0,1).$

\noindent Across various dimensions of the response variable and predictors in this heteroscedastic scenario, in Figure \ref{fig:figS3}, we demonstrate that the method maintains stable performance as the sample size increases and the number of predictors increase for a confidence level $\alpha= 0.95$. 

\begin{figure}[ht!]
\begin{center}
\begin{tabular}{ll}
\includegraphics[width=0.9\linewidth , height=.38\linewidth]{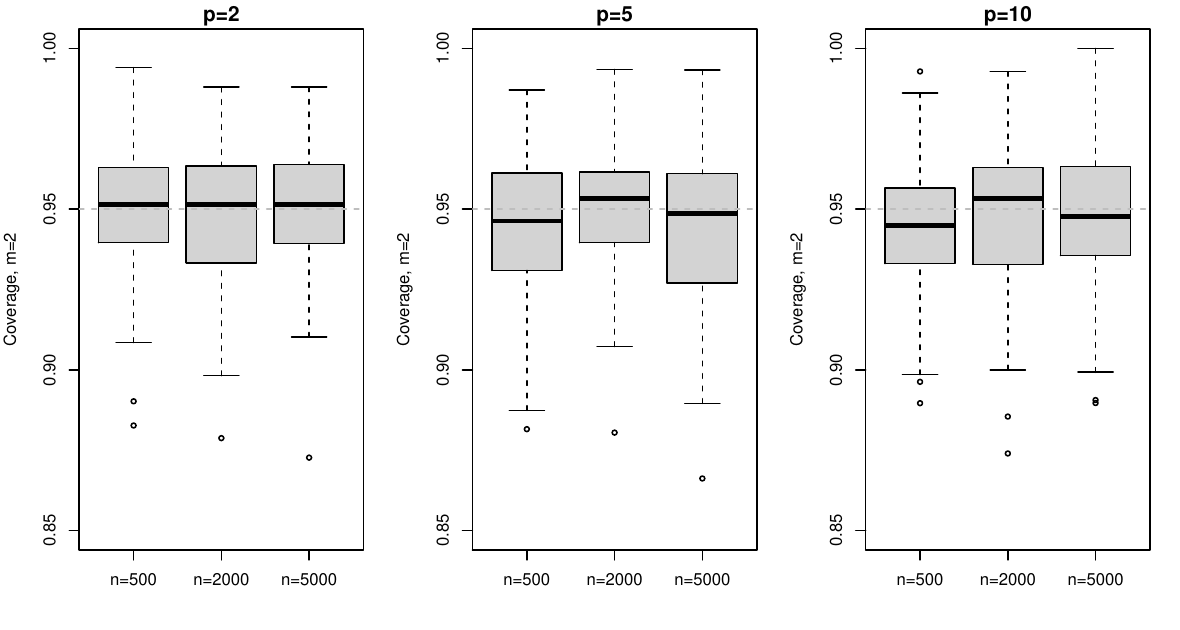} &
\end{tabular}
\end{center}
\caption{Estimated marginal coverage for multivariate data in the heteroscedastic regime for a bivariate Euclidean response.}
\label{fig:figS3}
\end{figure}

\subsection{Functional to Functional regression}



In order to show the versatility of our uncertainty quantification algorithm, we simulate a scenario with standard functional-to-functional models in the space $L^{2}([0,1]) \to  L^{2}([0,1])$. In particular, we suppose the following underlying regression model:

\begin{equation}
    Y(t) = \beta_0(t) + \int_{0}^{1} X(s) \beta(s,t)  \, ds + \epsilon(t),
\end{equation}

\noindent where $\beta_0(t) = \sin(\pi t)$, $\beta(s,t) = 5st$, $X(s) = s\eta$, $\eta\sim N(0,\sigma^{2}_{\eta})$ and $\epsilon(t) = \cos(2\pi t)\epsilon_1 + \sin(2\pi t)\epsilon_2$, and for $j \in \{1,2\}$, $\epsilon_{j} \sim N(0, \sigma_j^{2})$, where $\sigma_1=0.5,\sigma_2=0.75$. The covariate process $X_i(s)$ (i.i.d. copy of $X(s)$) is generated as $X_i(s)=\sum_{k=1}^{10}\psi_{ik}\phi_k(s)$, where $\phi_k(s)$ are orthogonal basis polynomials and $\psi_{ik}$ are  mean zero and independent normally distributed scores with variance $\sigma^2_k=(10-k+1)$.

Figure \ref{fig:figS3} (left) shows that in this functional-to-functional example, the non-asymptotic properties are maintained, as well as the statistical consistency.


\subsection{Probability distribution}
We assume a distributional regression model of the type 

\begin{equation}
    Q(p)=  10p+ X_1(1+p) +X_2(p^2) +\epsilon
\end{equation}
The covariates $X_1\sim Unif(0,1)$, $X_2\sim Unif(0,0.5)$ respectively. The residual error $\epsilon \sim N(0,0,9^2)$.

The boxplot in Figure \ref{fig:figS3} (right) shows that in this scalar-to-distributional regression model, the non-asymptotic properties are reached, as well as statistical consistency.

\begin{figure}[ht!]
\begin{center}
\begin{tabular}{ll}
\includegraphics[width=.5\linewidth , height=.5\linewidth]{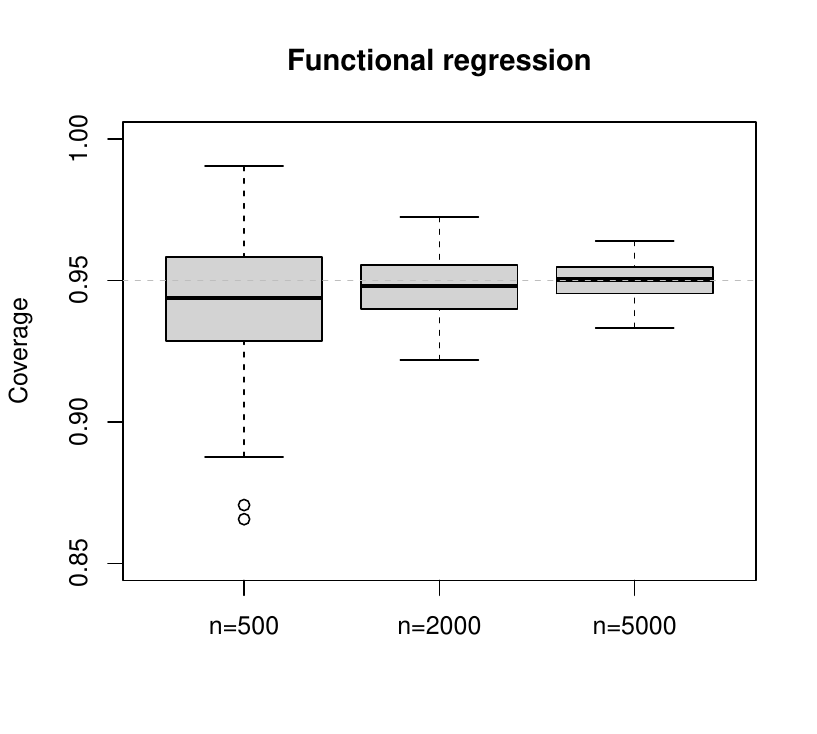} &
\includegraphics[width=.5\linewidth , height=.5\linewidth]{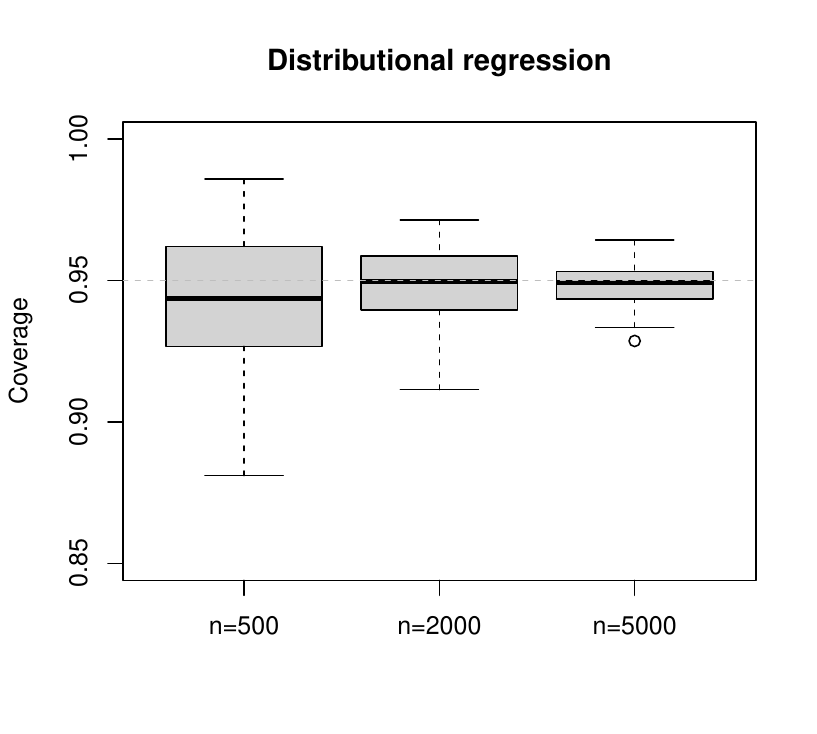}\\
\end{tabular}
\end{center}
\caption{Estimated marginal coverage, functional data (left) and distributional data (right), homoscedastic case, $\alpha=0.05$.}
\label{fig:figS3}
\end{figure}

	\section{Data example  application:  Physical activity recommendations with  NHANES accelerometer data}
	\subsection{Motivation and data description}
	
Physical exercise is considered one of the most effective pharmacological interventions across a broad spectrum of diseases, as well as a therapy for reversing physiological decline with age or improving overall health \cite{pedersen2006evidence,almeida2014150,bolin2018physical,mendes2016exercise,franks2017causal,friedenreich2021physical,burtscher2020run}. Although many medical guidelines suggest standard recommendations, such as performing $150$ minutes of aerobic exercise per week, there is a growing consensus in the community about the need for a personalized prescription, as well as patient assessment, necessary to ensure the success of the intervention performed \cite{mendes2016exercise,matabuena2019prediction,buford2013toward}. Nowadays, with the modern accelerometer and smartphones we have more reliable ways to characterize physical activity levels of a individual \cite{straczkiewicz2023one}.

Data modeling is a critical step in accelerometer data analysis for exploiting the rich information that accelerometers capture about physical activity patterns at different resolution time scales. However, as subjects are usually monitored in free-living conditions, practitioners usually resort to basic summary statistics, such as total expenditure normalized by the time the device is worn or compositional metrics. In our recent work \cite{10.1093/jrsssc/qlad007}, we have proposed a new functional distributional profile of the physical activity performed by the individual. The new representation automatically captures the previous summary metrics. Additionally, in the same paper, we generalize a series of nonparametric regression models to incorporate the complex survey design of the NHANES cohort and include kernel ridge regression methods to demonstrate the advantages of the new representation over traditional summary measures of accelerometer data.

The NHANES data provide new scientific opportunities for characterizing physical activity patterns across different age groups, ethnicity, or disease sub-types due to their robust and randomization study design in the American Population, unlike another important physical activity cohort such as US population. In this study, we select healthy individuals aged 18-80 years and exclude patients with specific comorbidities and other reasonable exclusion criteria explained in detail in the Supplementary Material.

The goal of this paper is to use the new conditional prediction regions algorithm to define, according to specific characteristics of the patients, such as age, sex, BMI, what are the expected distributional representation of the physical activity. Roughly speaking, we can define personalized physical activity recommendations for the individual of the US. population. In this way, it provides a  sophisticated tool to determine when physical activity  interventions  patterns must be carry out.

Given the increase in the use of wearable technology to monitor health status at the population level, the methods introduced here have the potential to contribute to the public health policies  on physical activity traditionally designed for the average person.



	\subsection{Modeling accelerometer data with distributional data}

	\subsubsection{Definition of functional representation}\label{section:representacion}
	
	We begin by presenting the formal definition of the accelerometer distributional representation, following the mathematical framework established in \cite{10.1093/jrsssc/qlad007}. Let $n_i$ be the number of observations for patient $i$, recorded as pairs $(t_{ij},X_{ij})$, $j=1,\cdots,n_i$. The $t_{ij}$ are time points in the interval $[0,T_i]$ at which the accelerometer records activity information, and $A_{ij}$ is the corresponding accelerometer measurement at time $t_{ij}$. We account for inactivity time by assigning a positive probability mass at zero, equal to the fraction of total time that the individual is physically inactive. However, the range of values measured by the accelerometer can vary widely between individuals and groups, presenting challenges when applying traditional compositional functional analysis methods (e.g., \cite{van2014bayes,petersen2016}).

To address these challenges, we propose using a cumulative distribution function $F_i(t)$ for each individual. Let $Y_i(t)$ be a latent process such that the accelerometer measures $A_{ij} = Y_i(t_{ij})$ $(j=1,\ldots,n_i)$, and define $F_i$ as
Let $Y_i(t)$ be a latent process such that the accelerometer measures $A_{ij} = Y_i(t_{ij})$ $(j=1,\ldots,n_i)$, and define $F_i$ as

\begin{equation}\label{rep1}
F_i(x) = \frac{1}{T_i} \int_{0}^{T_i} \mathbf{1}\{Y_i(t) \leq x\} \, dt, \quad \text{for } x \geq 0.
\end{equation}

In this paper, we consider an invariant positive kernel $k: \mathcal{Y}\times \mathcal{Y}\to \mathbb{R}^{+}$ equipped with the $2$-Wassertein distance. Specifically, we define $k(F,G)=\kappa(d_{W^{2}}\left(F,G\right))$, where $\kappa(\sqrt{\cdot})$ is a completely monotone, positive function. In our data analysis examples, we use an explicit $2$-Wassertein Gaussian kernel defined as

\begin{equation}
k(F,G)= e^{ \sigma \int_{0}^{1} (F^{-1}(s)-G^{-1}(s))^{2} \mathrm{d}s},
\end{equation}
where $\sigma>0$ is the kernel bandwidth. Importantly, to carry out the statistical modeling, we must estimate $F_{i}^{-1}$ from the random sample $\{A_{ij}\}^{n_i}_{j=1}$. In practice, we use the empirical estimator $\widehat{F}_{i}(t)= \frac{1}{n_i} \sum_{j=1}^{n_i} 1\{A_{ij}\leq t \}$, $t\in \mathbb{R}$.
	
\begin{figure}[h]
    \centering
    \begin{subfigure}[b]{0.4\textwidth}
        \includegraphics[width=1\textwidth]{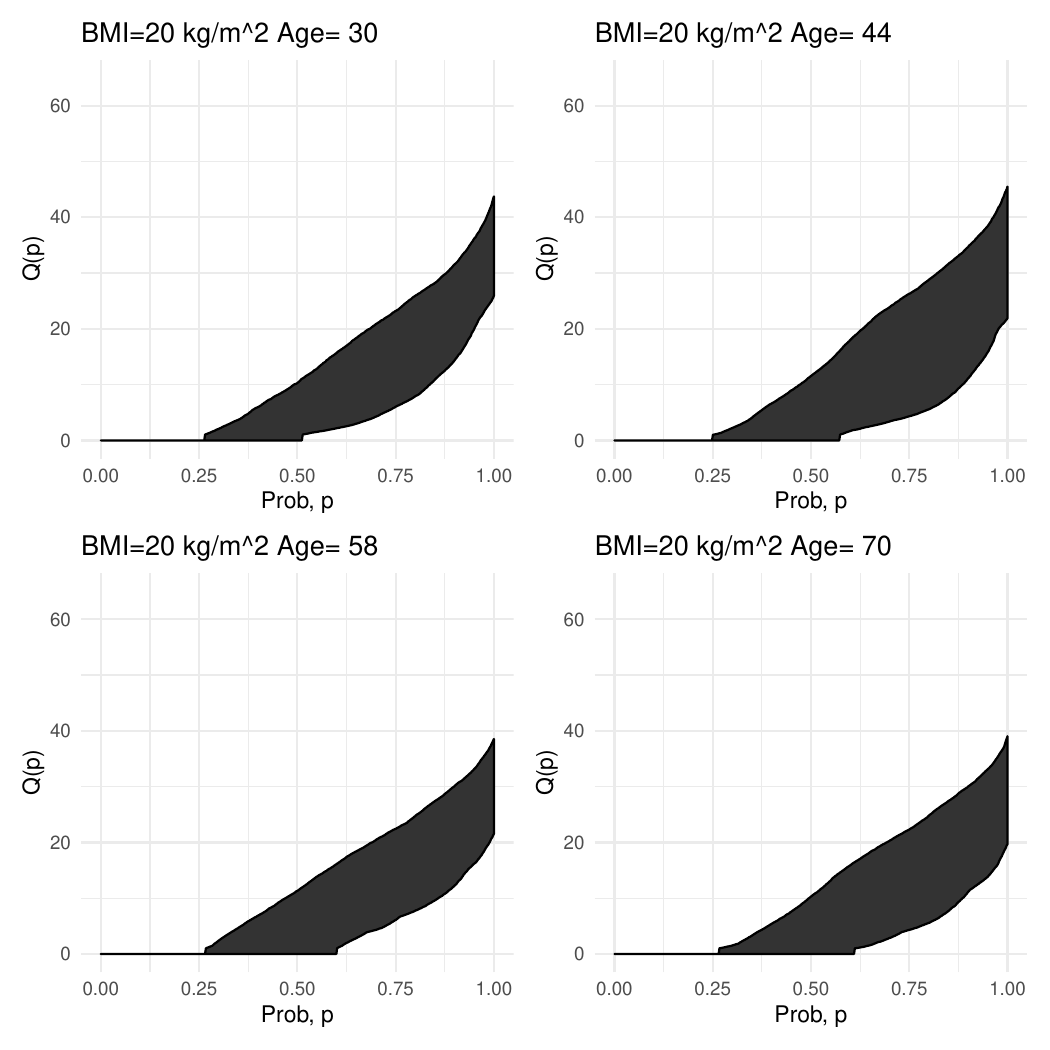} 
        \caption{Women, BMI= $20kg/m^{2}$}
        \label{fig:figura1}
    \end{subfigure}
    
    \begin{subfigure}[b]{0.4\textwidth}
        \includegraphics[width=1\textwidth]{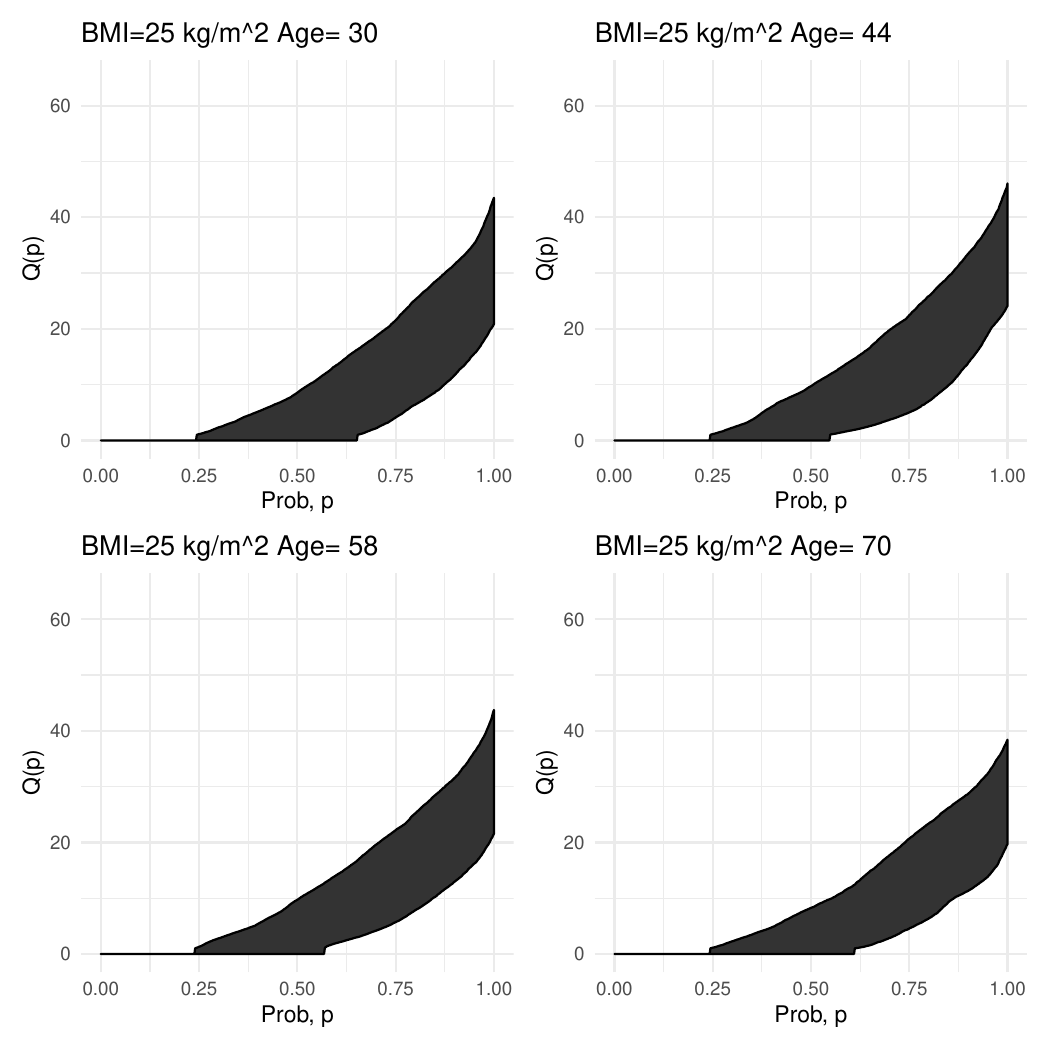} 
        \caption{Women, BMI= $25kg/m^{2}$}
        \label{fig:figura2}
    \end{subfigure}
    
    \begin{subfigure}[b]{0.4\textwidth}
        \includegraphics[width=1\textwidth]{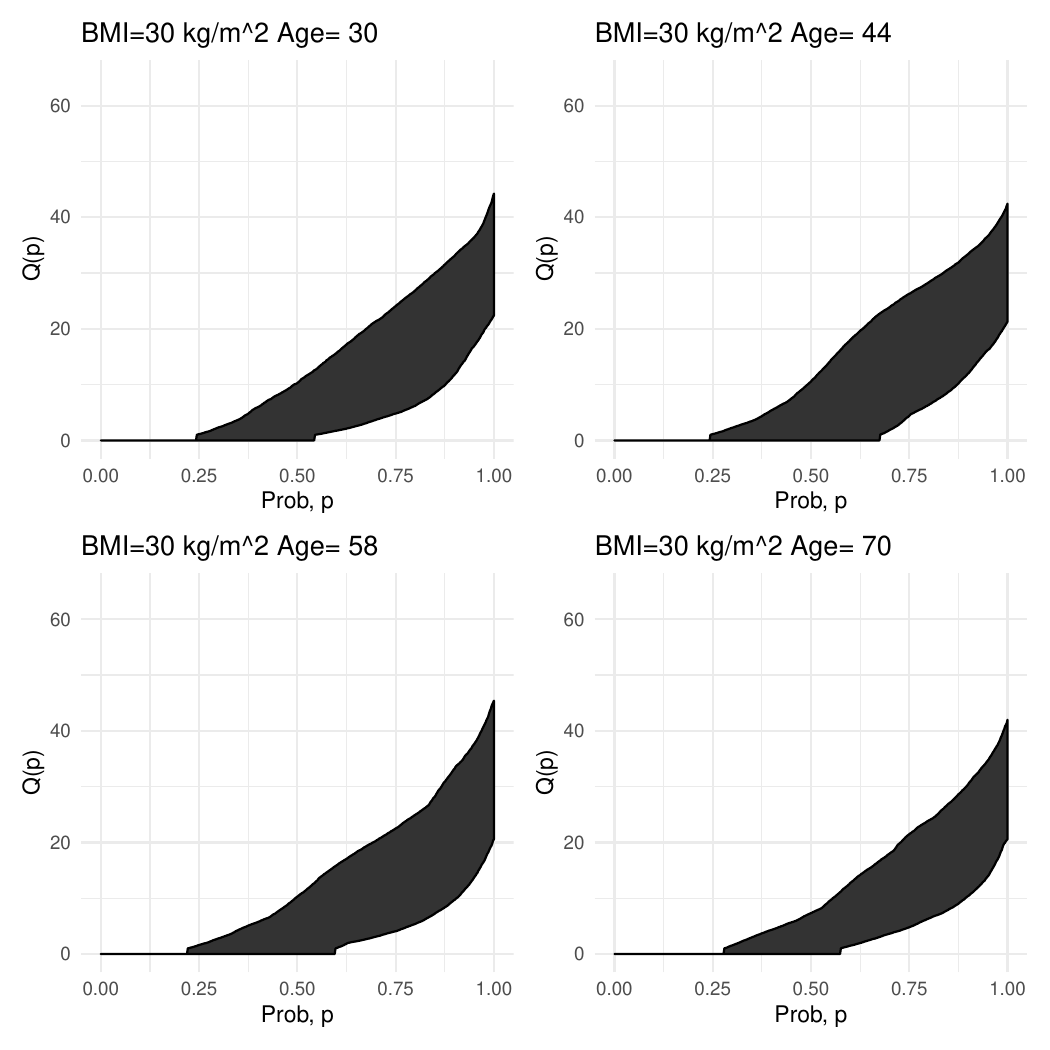} 
        \caption{Women, BMI= $30kg/m^{2}$}
        \label{fig:figura3}
    \end{subfigure}
    
    \caption{Expected Quantile of Physical Activity Across Different Ages and Body Masses in Healthy Women of the American Population for $\alpha=0.5$.}
    \label{fig:tres_figuras}
\end{figure}

\subsection{Conditional kernel mean embedding for survey designs}
In the NHANES database, we observe the set $\{(X_{i},Y_{i},w_{i})\}_{i=1}^{n}$, where for the $i^{\text{th}}$ observation, we have available the $i^{\text{th}}$ survey weights. To accommodate this information about survey weights, we must modify the related mean square problem as:

\begin{equation*}
\underset{C: \mathcal{H}_{\mathcal{X}} \rightarrow \mathcal{H}_{\mathcal{Y}}}{\arg \min }\frac{1}{n}\sum_{i=1}^{n}w_{i} \left\|\phi\left(y_{i}\right)-C \psi\left(x_{i}\right)\right\|_{\mathcal{H}_{\mathcal{Y}}}^{2}+\lambda\|C\|_{\mathrm{HS}}^{2},
\end{equation*}

\noindent where $\|C\|_{\mathrm{HS}}$ denotes the Hilbert-Schmidt norm. Finally, we arrive at the final empirical estimator as:

\begin{equation*}
\hat{\mu}_{P(Y \mid X)}(x) = \hat{C} \psi(x) = \sum_{i=1}^{n} \beta_{i}(x) \phi\left(y_{i}\right) = \Phi \boldsymbol{\beta}(x).
\end{equation*}

\subsection{Results}

\subsubsection{Recommendation physical values in different age group and body mass compositions}

We accommodate survey weights in the conformal algorithm and provide several predictions of expected physical activity values with a confidence level of \(\alpha=0.5\), adjusted by sex, age, and body mass index (BMI). For this purpose, we only include individuals without chronic diseases in the analysis, focusing specifically on female participants. Figure \ref{fig:tres_figuras} shows the results for different combinations of BMI and age. Generally, we observe a decline in the values within the prediction regions as BMI and age increase. The results highlight the importance of considering physical activity levels according to the specific characteristics of individuals.

\subsubsection{Estimation of quality of physical activity performed}

Our next goal is to define a ranking for the level of outlier among these observations. For each individual $(X_{i}, Y_{i}), i \in \mathcal{D}_{n}$, we define the variable $\alpha_{i}$ as the minimun $\alpha \in (0,1)$ for which the indicator function $\mathbb{I}\{Y_{i} \in \widehat{C}^{\alpha}(X_i)\} = 1$. Focusing on the three patients in Figure \ref{fig:tres}, presented in the introduction of the paper, we can infer that the $\alpha_{i}$ values are 96\%, $1\%$ and 94\%, respectively, indicating that with a high probability, patients $a$ and $c$ are outliers, while the patient b is in the median of the conditional distribution. Given that the Y-axis indicates that higher values are more positive, we deduce that the first patient is an outlier for being very active, while the second is sedentary. An important advantage of our algorithm is that it provides global prediction regions, and for this purpose, we do not need to resort to the geometry of the supremum norm. The new method provide a formal criteria to detect anomalous behavior of the patient physical activity trajectories. The method can be local in each probability of quantile function $p\in [0,1]$, and for global perspective across $p$.

\section{Discussion}
This paper introduces a novel framework for uncertainty quantification in random statistical objects defined in separable Hilbert spaces, utilizing conformal prediction techniques and the depth-measure derived from a 
kernel mean embedding. The efficacy of the new algorithms with finite--sample sizes has been thoroughly validated through an extensive simulation study, covering a spectrum of situations that include multivariate Euclidean data and functional responses. We also illustrate the algorithms in a physical activity application where functional information from accelerometer devices has been considered. The analysis of applications accentuates the necessity of creating specific individual decision-making to personalize physical guidelines according to patient characteristics.
Although the exposition within this manuscript predominantly focuses on the utilization of functional data, it is crucial to acknowledge the broader potential applicability of the methodology introduced to various complex data structures, including but not limited to graphs and trees, or any random object embeddings within a separable Hilbert space \cite{schoenberg1937certain,schoenberg1938metric}. The methods proposed here can be combined with any depth-measure and not only with a kernel mean embedding  (see for example \cite{hallin2021distribution}).
In a future work,  we focus on the extension of this kernel uncertainity quantification framework as the case of time-to event data \cite{garcia2023causal}
 and dependent data.
%


\end{document}